\documentclass[structabstract]{aa}  
\usepackage{graphicx}
\usepackage{lscape}
\usepackage{longtable}
\usepackage{enumitem}
\usepackage{float}
\usepackage{caption}

\usepackage{txfonts}
\usepackage{soul}

\def\mstar  {$M_{*}$}
\def\macc   {$\dot{M}_{\rm acc}$}
\def\mout   {$\dot{M}_{\rm out}$}
\def\lacc   {$L_{\rm acc}$}

\def\kms  {km s$^{-1}$}
\def\msun {$M_{\odot}$}
\def\lsun {$L_{\odot}$}
\def\lstar {$L_*$}

\def\henir {He I $\lambda$10830~\r{A} }

\usepackage{xcolor}

\begin{document}
   \title{The \henir line as a probe of winds and accretion \\in young stars in Lupus and Upper Scorpius \thanks{This work is based on observations made with ESO telescopes at the La Silla Paranal Observatory under programme ID  084.C-0269, 084.C-1095, 086.C-0173, 087.C-0244, 089.C-0143, 085.C-0238, 085.C-0764, 089.C-0840, 090.C-0253, 093.C-0658, 094.C-0913, 095.C-0134, 095.C-0378, 097.C-0349, 097.C-0378, and 0101.C-0866, and archive data of programmes 085.C-0764 and 093.C-0506.}} 

   \author{J. Erkal\inst{\ref{instESO},\ref{instUCD}}, C.F. Manara\inst{\ref{instESO}}, P.C. Schneider\inst{\ref{instHamburg}}, M. Vincenzi\inst{\ref{instDuke}}, B. Nisini\inst{\ref{instRome}}, D. Coffey\inst{\ref{instUCD}}, J.M. Alcal\'a\inst{\ref{instNapoli}}, D. Fedele\inst{\ref{instFirenze}}, S. Antoniucci \inst{\ref{instRome}}
          }
          
     \authorrunning{J. Erkal et al.}
     \titlerunning{HeI line as a probe of winds and accretion}

 \institute{European Southern Observatory, Karl-Schwarzschild-Strasse 2, 85748 Garching bei M\"unchen, Germany\label{instESO}\\  \email{jessica.erkal@eso.org}
\and
School of Physics, University College Dublin, Belfield, Dublin 4, Ireland\label{instUCD}
\and
Hamburger Sternwarte, Gojenbergsweg 112, 21029, Hamburg, Germany\label{instHamburg}
\and
Department of Physics, Duke University, Durham, NC 27708, USA\label{instDuke}
\and
INAF-Osservatorio Astronomico di Roma, Via di Frascati 33, 00078 Monte Porzio Catone, Italy\label{instRome}
\and
INAF-Osservatorio Astronomico di Capodimonte, via Moiariello 16, 80131 Napoli, Italy\label{instNapoli}
\and 
INAF-Osservatorio Astrofisico di Arcetri, Largo Enrico Fermi 5, 50125, Firenze, Italy\label{instFirenze}
}


 
  \abstract
   {The \henir line is a high excitation line which allows us to probe the material in the innermost regions of protostellar disks, and to trace both accreting and outflowing material simultaneously.}
   {We use X-Shooter observations of a sample of 107 young stars in the Lupus (1--3 Myr) and Upper Scorpius (5--10 Myr) star-forming regions to search for correlations between the line properties, as well as the disk inclination and accretion luminosity.}
   {We identified eight distinct profile types in the sample. We fitted Gaussian curves to the absorption and/or emission features in the line to measure the maximum velocities traced in absorption, the full-width half-maximum (FWHM) of the line features, and the Gaussian area of the features.}
   {We compare the proportion of each profile type in our sample to previous studies in Taurus. We find significant variations between Taurus and Lupus in the proportion of P Cygni and inverse P Cygni profiles, and between Lupus and Upper Scorpius in the number of emission-only and combination profile types. We examine the emission-only profiles in our sample individually and find that most sources (nine out of 12) with emission-only profiles are associated with known jets. When examining the absorption features, we find that the blue-shifted absorption features appear less blue-shifted at disk inclinations close to edge-on, which is in line with past works, but no such trend with inclination is observed in the sources with only red-shifted features. Additionally, we do not see a strong correlation between the FWHM and inclination. Higher accretion rates were observed in sources with strong blue-shifted features which, along with the changes in the proportions of each profile type observed in the two regions, indicates that younger sources may drive stronger jets or winds.}
  {Overall, we observe variations in the proportion of each \henir profile type and in the line properties which indicates an evolution of accretion and ejection signatures over time, and with source properties. These results confirm past works and models of the \henir line, but for a larger sample and for multiple star-forming regions. This work highlights the power of the \henir line as a probe of the gas in the innermost regions of the disk.}

   \keywords{Stars: pre-main sequence -- stars: formation -- protoplanetary disks -- accretion, accretion disks  }

 \authorrunning{Erkal et al.}
\titlerunning{Helium lines}
\maketitle
%

\section{Introduction}

From the earliest stages of star formation, jets and outflows are commonly observed from young stars \citep{frank14}. At the same time, accretion occurs as material flows radially inwards through the disk onto the star \citep{hartmann16}. The interplay between these two processes is relevant for the final stellar mass and for the removal of disk mass, which impacts the material available for planet formation \citep{manara18b}. However, the connection between accretion and ejection is not fully understood \citep[e.g.][]{pascucci2022,manara22}.

Accretion occurs as material is transported through the disk. When the material reaches the disk truncation radius (R$_{t}$ $\simeq$ 3-7~R$_{\star}$), it falls along magnetic field lines in accretion flows onto the star \citep{hartmann16}. As the material falls onto the star, accretion shocks lead to excess emission in ultraviolet and optical wavelengths, which causes photospheric absorption features in the spectra to appear shallower (i.e. veiling). Veiling also occurs at infrared (IR) wavelengths (e.g. \citealt{folha1999}, \citealt{fischer11}).

Magnetic fields in the disk are likely responsible for the launching and collimation of jets and outflows seen in young stars \citep{pudritzray2019}. The highly collimated, high velocity jets are generally observed to have a knotty structure, suggesting that material is not ejected at a constant rate. If these jets are accretion powered, the knotty structure points to variable accretion with periods of high accretion in accretion bursts and a smaller contribution made by a lower constant accretion rate (e.g. \citealt{arce07,vorobyov2018}). Often surrounding the high-velocity jet is a slower wind ejected directly from the disk (photoevaporative winds or magnetohydrodynamic (MHD) winds, for example, \citealp{pudritz83, EP17}). In younger Class 0/I sources, these outflows may also contain entrained material swept up by the jet \citep[e.g.][]{zhang2019}.
These jets and outflows can be traced back to the star-disk plane, and they appear to be launched through MHD processes from the inner regions of the disk \citep{bp82,pudritz83}, though the exact launching mechanism is not yet known \citep{ferreira2006}. However, it is still not clear which mechanism dominates the mass removal throughout the evolution of the protostar, as photoevaporative winds \citep[see e.g.][]{EP17} may also drive the wider outflows. Photoevaporation is important for disk dispersal, but it does not contribute to the removal of angular momentum. The launching of a magnetically driven jet from the disk is believed to play a role in angular momentum removal, which drives accretion onto the star \citep{EP17,bai16}, thus highlighting the connection between accretion and ejection in young stellar systems. This is further supported by the correlation between mass loss rates and the disk accretion rate (\mout/\macc $\simeq$ 10\%, e.g. \citealt{cabrit90}, \citealt{hartigan95}, \citealt{nisini18}). On the contrary, through viscous evolution, angular momentum is transported through the disk \citep{hartmann98}, rather than being removed via an outflow, and so it is not possible to find a connection between accretion and ejection in this scenario. An exception to this is in the case of X-winds which are launched near the corotation radius, where the magnetosphere truncates the disk (see e.g. \citealt{shu1994}). Many of these winds and outflows are traced by forbidden emission lines (e.g. [\ion{O}{I}], [\ion{S}{II}], [\ion{Fe}{II}]; see e.g. \citealt{pascucci2022}) within which a high-velocity component (HVC) and low-velocity component (LVC) are often detected. The HVC is attributed to the high-velocity jets, while the LVC is linked to the slower disk winds which show correlations with the accretion luminosity (e.g. in the [\ion{O}{I}]~$\lambda$6300~\r{A}) line. However, to understand the true nature of the connection between accretion and ejection in young stars, we must find a way to examine the innermost regions of the disk.

High excitation lines have been used as a probe of the inner regions of the disk, since their formation is confined to high temperature regions or near ionizing radiation \citep{beristain01,edwards03}, in the inner au \citep{edwards09}. One such line is the \henir line. With high excitation and a metastable lower level, this line is particularly sensitive to sub-continuum absorption providing a way to probe both accretion and ejection in the inner disk regions simultaneously. 
Red-shifted absorption in this line traces accreting material, travelling near free-fall velocity onto the star. The red-shifted absorption creates the inverse P Cygni profile and is generally seen at velocities between 0 to 350~\kms\, \citep{fischer2008}. Red-shifted absorption was identified as a good tracer of accretion as \citet{edwards06} noted that even though this feature is often observed at Classical T Tauri stars (CTTSs) with ongoing accretion, no such feature is observed at non-accreting Weak-line T Tauri stars (WTTSs; see Figure \ref{fig:classIII} for \henir line profiles in non-accreting Class III sources). However, the absence of red-shifted absorption
may point to the presence of emission from a wind which fills the absorption feature, or high veiling of the star. \citet{thanathibodee2022}, for example, used the \henir line to study young stars previously thought to be non-accretors. 

Winds are traced by blue-shifted absorption which traces outflowing material, and can be linked to a stellar wind or disk wind depending on the absorption profile. Wide blue-shifted absorption features are formed by a radially expanding stellar wind which is traced up to several hundred \kms, whereas narrow low-velocity absorption features indicate the presence of a disk wind, where only a small range of velocities in the wind are intercepted along the line of sight to the star \citep{edwards06,edwards09}. Emission may also trace a polar wind at the star if the star is highly inclined \citep{kwan2007}. 
The \henir line likely traces a stellar wind because narrow absorption features in the line, characteristic of disk winds, are observed less frequently than stellar wind signatures \citep{edwards06,kwan2007}. Nonetheless, both a disk wind and stellar wind could coexist at these stars, both removing mass and angular momentum from the system. Red-shifted absorption, tracing accreting gas, was observed more frequently in the He I line than in other lines (e.g. Pa$\gamma$) where veiling likely prevents the detection of this feature as shown by \citet{fischer2008}. 

However, past studies have only dealt with small samples (n $\approx$ 30) of stars of the same age, i.e. all situated in the same star forming region, e.g. Taurus \citep[1-3 Myr,][]{Krolikowski21}. Observing larger samples of stars from different star forming regions is key to establishing a connection between the accretion and ejection in these sources, for various stages of evolution. 

In this paper we present the analysis of a sample of over 100 young stars in the Lupus \citep[1--3 Myr,][]{comeron08,luhman20} and Upper Scorpius \citep[5--10 Myr,][]{PM12,luhman22} star forming regions. We observe the \henir line at higher resolution and for a larger sample size than past works, providing an opportunity to identify statistically significant trends in the data. The \henir line is examined for each star and categorised into profile types to search for trends with respect to age, stellar properties and accretion properties. 

The paper is organised as follows. The sample, observations and data reduction are presented in Section \ref{sect::obs}. In Section 3 we describe the method used for identifying the \henir profile types (Section \ref{sect:fitting}). Here we also describe each profile type identified in our sample (Section \ref{sect:profile_types}). In Section \ref{sect::Res} we present statistics on profile type occurrences in each region and measured absorption velocities. We discuss our results in Section \ref{sect::disc}, focusing in particular on observed trends in velocities, and dependences on source inclination and accretion properties. Our main conclusions are summarised in Section \ref{sect::concl}.


\section{Sample, observations, and data reduction}
\label{sect::obs}

\subsection{Sample}

\begin{figure}[h]
    \centering
    \includegraphics[width=0.45\textwidth]{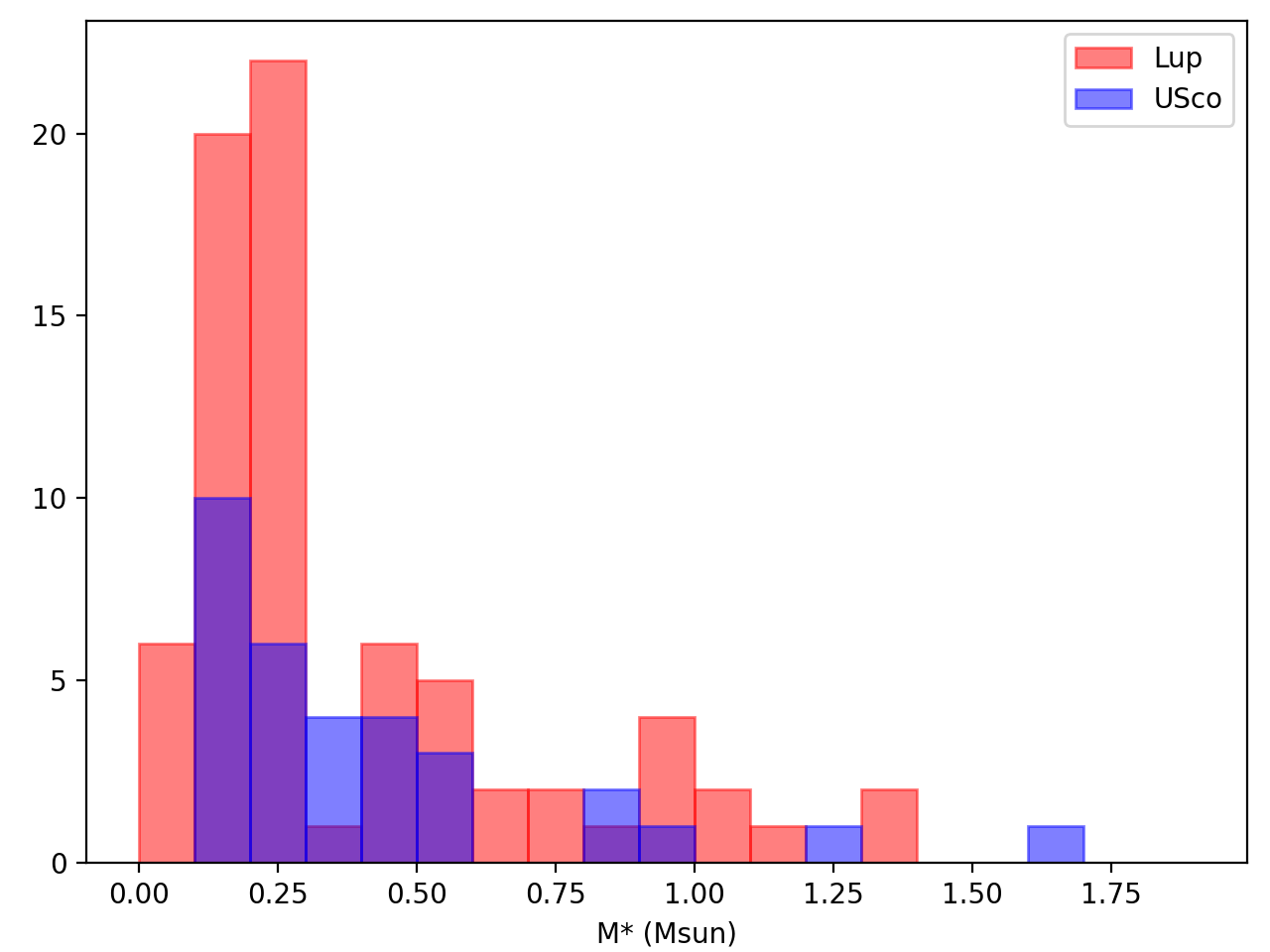}
    \caption{Stellar masses observed in both the Lupus (red) and Upper Scorpius (blue) regions.}
    \label{fig:hist_mstar}
\end{figure}

\begin{figure}[h]
    \centering
    \includegraphics[width=0.45\textwidth]{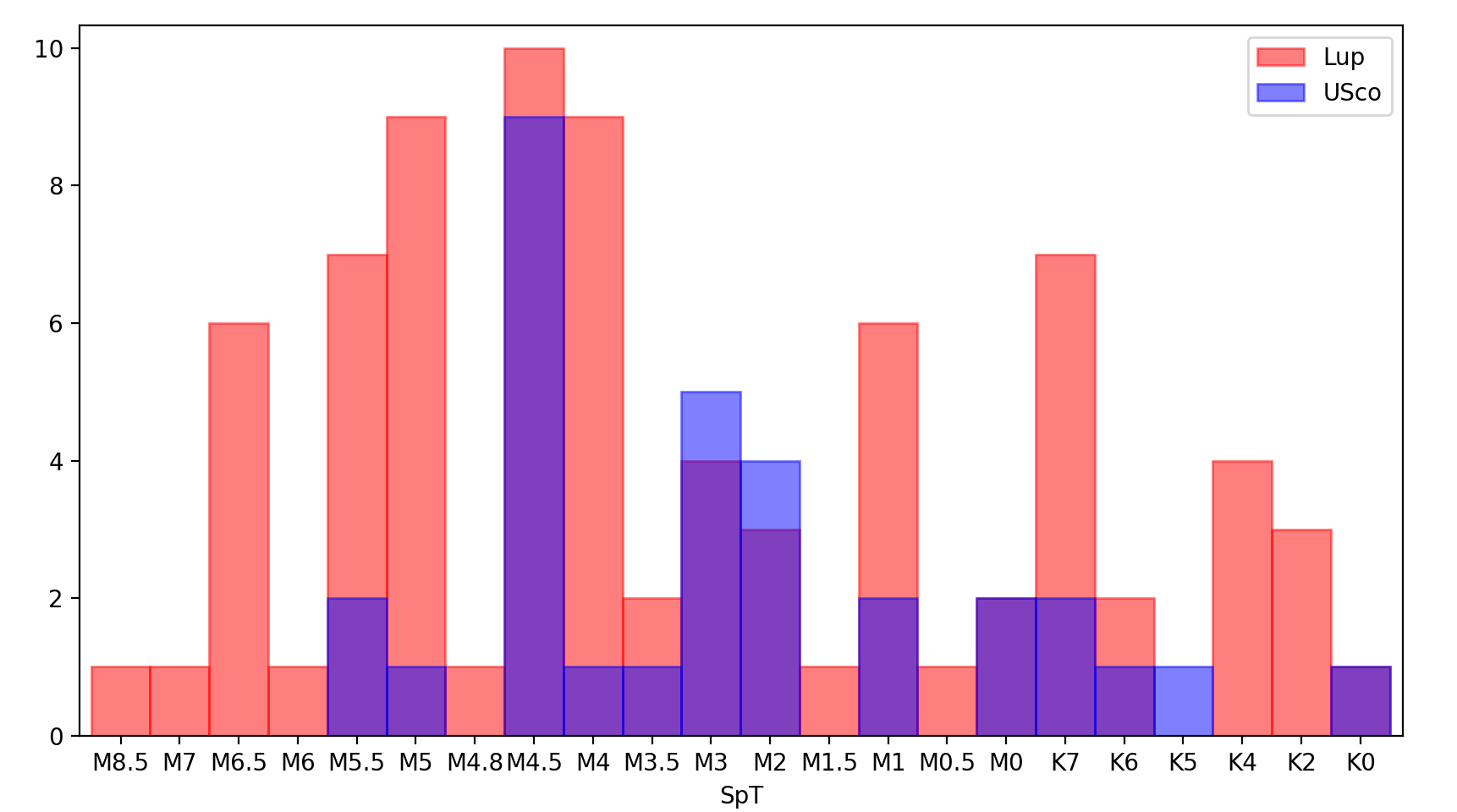}
    \caption{Spectral types of the central star observed in both the Lupus (red) and Upper Scorpius (blue) regions.}
    \label{fig:hist_spt}
\end{figure}

Our sample consists of a total of 117 young stellar objects - 82 Class II sources are located in the Lupus star forming region (d $\simeq$ 160 pc; 
\citealt{alcala14,alcala17}) and a further 35 sources in Upper Scorpius (d $\simeq$ 145 pc; 
\citealt{manara20}). The Lupus sample is younger (1 -- 3 Myr; \citealt{comeron08}) compared to the Upper Scorpius sample (5 -- 10 Myr; \citealt{luhman2012}), allowing us to investigate the evolution of the \henir profile for different ages. The targets in our sample have spectral types between K0 and M8.5, with stellar masses in the range of \mstar $\simeq$ 0.1 - 1.6 \msun (see Figures \ref{fig:hist_mstar} and \ref{fig:hist_spt}), allowing us to investigate the properties of the \henir line in relation to varying stellar properties such as spectral type, accretion luminosity, among others. We include information on the spectral types, disk inclination, and accretion properties for the individual targets in Tables A.1 and A.2, 
however the sample is described in depth in \citet{alcala14,alcala17} and \citet{manara20}. 

\subsection{Observations}
All the observations included in this work have been obtained with the ESO/VLT X-Shooter spectrograph \citep{vernet11}. This medium-resolution and high-sensitivity instrument simultaneously covers the wavelength range between $\sim$ 300 nm and $\sim$ 2500 nm. The spectra are divided in three arms which cover three different wavelength ranges. The helium line analysed here is located 
in the NIR arm, with resolution R $\sim$10500-5300 depending on the slit width. The spectra used in this analysis have been reduced with the procedure described by \citet{alcala14}, and the observational details are listed in \citet{alcala14,alcala17} and \citet{manara20} depending on the targets, as discussed in the following section.

\subsection{Data reduction}
The data were initially reduced with the X-Shooter pipeline \citep{modigliani10} using the ESO Reflex workflow. 
This process is described in detail in previous works where the same data reduction steps are taken, for example \cite{alcala17} and \cite{manara20}. The telluric correction procedures for both regions are also described in these papers. Telluric correction for the Lupus sample was performed using two separate methods on the NIR and VIS arms \citep[see Appendix A in][ for detail]{alcala14}. The telluric correction for Upper Sco sources was carried out using molecfit \citep{smette2015,kausch2015} as described by \citet{manara20}. Radial velocity (from \citealt{frasca2017,manara20}) and heliocentric velocity corrections are applied to both the NIR and VIS spectra. To correct for wavelength shifts between the two arms, we used a 1D Gaussian fit to measure the centroid of the Lithium 670.78~nm absorption line and the Pa$\delta$ 1004.94~nm line (observed in both the NIR and VIS spectra). We first fitted the Li 670.78~nm line with a 1D Gaussian, finding median Lithium shifts of $\approx$-0.5~\kms\, and -0.4~\kms\, for Lupus and Upper Sco. The Li shift is applied to both NIR and VIS spectra, such that the lithium line is centred on 0~\kms. We then corrected for wavelength shifts between the NIR and VIS spectra using the Pa$\delta$ line present in both arms. We found median Pa$\delta$ shifts of $\approx$ 9~\kms\, and -2.96 \kms\, for the Lupus and Upper Sco samples (respectively), which is then applied to the NIR spectrum.  
In a few cases where the signal-to-noise was too low for the lines to be properly fitted, we applied the median shifts listed above to these spectra. We find typical 1$\sigma$ errors on the Li and Pa$\delta$ velocity corrections of 2.96~\kms and 2.24~\kms, respectively. Adding these in quadrature, the combined error due to the velocity corrections is 3.7~\kms and thus do not introduce large uncertainties in the observed maximum velocities in the \henir line (see Section \ref{sect::Res}). We describe the process of aligning the NIR and VIS spectra in more detail in Appendix \ref{sect::aligning}.

\subsection{Ancillary data}
We have gathered further information on our sample for the analyses presented later in this paper. We use the stellar mass and luminosities, and accretion properties ($\dot{M}_{\rm acc}$ and L$_{\rm acc}$) from \citet{alcala19} for the Lupus sources, and from \citet{manara20} for Upper Scorpius sources, with distances for the individual sources from DR2 of the \citet{GaiaCollaboration2018}. The disk inclinations are known for 103 (88\%) of the 117 sources in our sample (85\% in Lupus, 94\% in Upper Sco) and were measured from ALMA continuum observations for the Lupus sample by \citet{tazzari17}, \citet{ansdell18}, and \citet{yen2018} and for Upper Scorpius by \citet{Barenfeld2017}. We note that these are measurements of the outer disk inclinations, whereas the \henir line traces gas close to the star. Since it is possible that there is a misalignment between the inner and outer disk \citep[e.g.][]{bohn2021} this possible issue will be considered when the results are analysed.

\section{Data analysis}

\subsection{Automatic determination of line profile types}
\label{sect:fitting}

We examined the \henir line profile in all of our targets using \textit{scipy} routines \citep{2020SciPy} in Python 3. We located the positions of all minima and maxima in the spectra between $\pm$ 400~\kms with the \textit{find\_peaks} routine, considering only those extrema with absolute intensities larger than twice the rms to avoid measuring noise. Using the location (in velocity) of the extrema and their height relative to the continuum (i.e. absorption or emission), we can distinguish between various profile types, described in further detail in Section \ref{sect:profile_types}. We used this initial information on the extrema and profile types to provide initial guesses on the peak parameters (amplitude, centroid in \kms and standard deviation in \kms) for Gaussian fitting. Depending on the profile type, between one to four 1D Gaussians were fitted to the \henir profiles using \textit{scipy}'s \textit{curve\_fit} routine, by adding the suitable number of 1D Gaussian functions defined as:
\begin{equation}
\centering
    f(x) = a e^{-\frac{1}{2}(\frac{x-\mu}{\sigma})^{2}}
\end{equation}

where \textit{a} is the amplitude, $\mu$ is the centroid and $\sigma$ is the standard deviation of the Gaussian peak.

We measured the $\chi ^{2}$ to identify poor fits (i.e. fits with $\chi ^{2}$ > 0.5) which we fitted again manually using new initial guesses for the peak parameters and profile type. We also visually checked each fit to ensure the profile is well-fitted (see Figures \ref{fig:Lup1}, \ref{fig:Lup2} and \ref{fig:USco} for each fitted profile). Out of the 82 sources in Lupus, 30 sources could not be automatically fitted and required manual inputs for a good fit. These profiles generally were categorised into the low-velocity absorption or miscellaneous profile types. 
Additionally, some of the profile types for these sources were identified incorrectly, therefore they required manual inputs and bounds to accurately fit the \henir line profile. These sources are listed in Table \ref{tab:lup_results1}.
Only one \henir profile in Lupus (2MASSJ16085373-3914367) could not be fitted due to low signal-to-noise and/or bad pixels. In Upper Scorpius, all sources could be fitted using a combination of 1D Gaussians, with 15 sources requiring manual inputs to be accurately fitted - these are listed in Table \ref{tab:usco_results1}.

\subsection{\henir line profiles}
\label{sect:profile_types}

\begin{figure}[ht]
    \centering
    \includegraphics[width=0.5\textwidth]{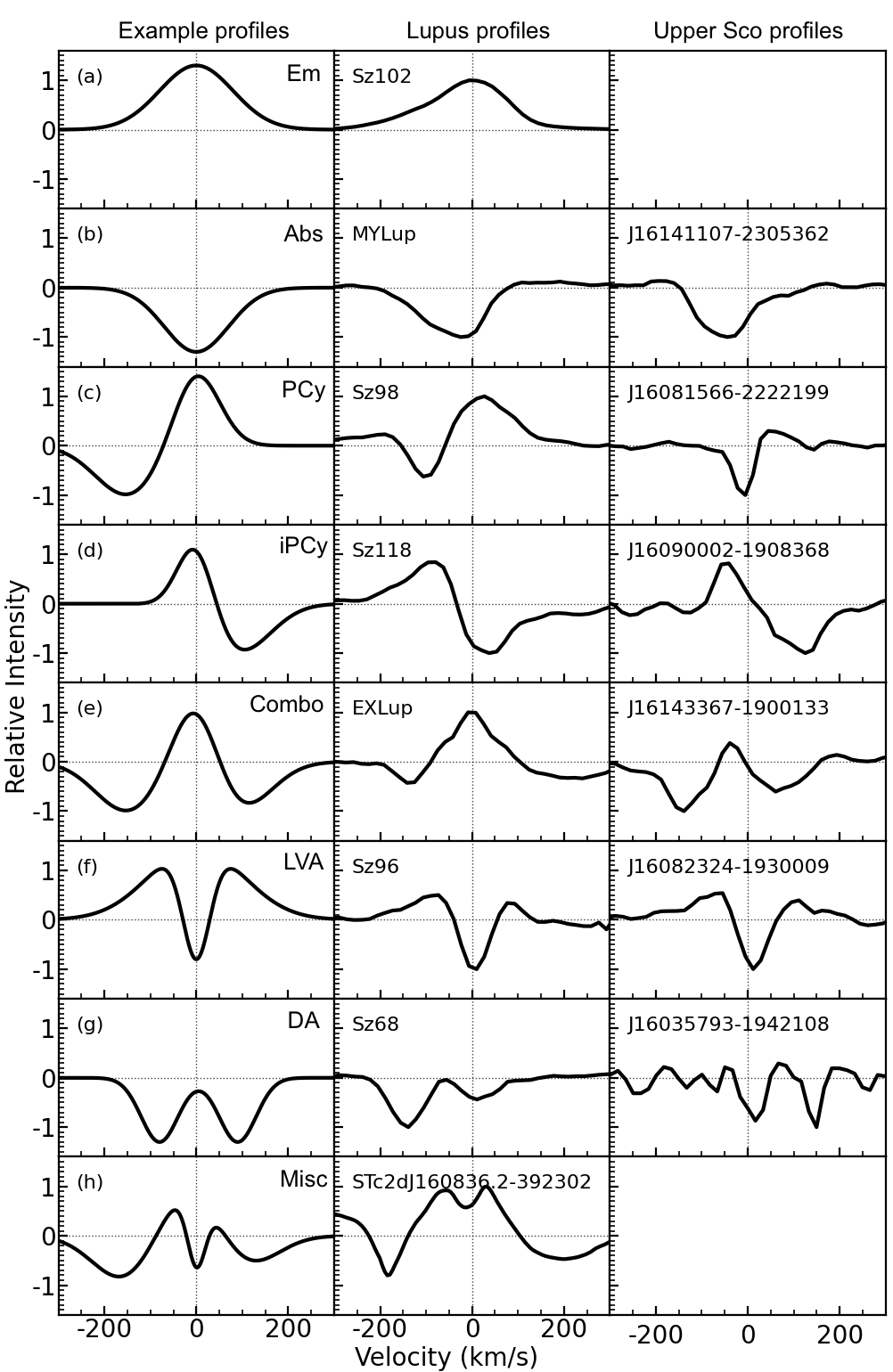}
    \caption{Theoretical profile types and example \henir profiles observed in our data. The left column shows the theoretical profiles for the eight profile types described in Section \ref{sect:profile_types}. The middle and right columns show examples of each profile type in the Lupus and Upper Scorpius samples, respectively. The flux has been normalised to the continuum level in each source, and rescaled between $\pm$ 1 for plotting. The grey horizontal line marks the continuum level at 0, while the vertical grey line denotes 0~\kms. We observe no pure emission or miscellaneous profile types in the Upper Scorpius sample. }
    \label{fig:profile_types}
\end{figure}

We identified eight distinct profile types in the data. The conceptual line profiles are presented in the left column of Figure \ref{fig:profile_types} and were created by adding between one to four one-dimensional Gaussians to the continuum, which is set at zero for all profiles. The profile types are as follows:

\begin{enumerate}[label=(\alph*)]

\item{Pure emission profiles are characterised by a single peak in emission. There are no sub-continuum absorption features present in these profiles.} 

\item{Pure absorption profiles are characterised by a single absorption feature. There are no other absorption features, or emission observed in these profiles.}

\item{P-Cygni profiles are characterised by the presence of a sub-continuum blue-shifted absorption feature, and emission centred near 0~\kms or slightly red-shifted. There is no red-shifted absorption feature observed.}

\item{Inverse P-Cygni profiles are characterised by the presence of a sub-continuum red-shifted absorption feature, and emission centred at low velocities. There is no blue-shifted absorption feature observed. The red-shifted absorption feature is generally wider than the absorption observed in P-Cygni profile types.}

\item{Combination profiles are characterised by the presence of two absorption features, one is red-shifted and the other is blue-shifted, and emission centred near $\sim$ 0~\kms. These profiles are a combination of the P-Cygni and Inverse P-Cygni profiles, with similar profile characteristics in both the blue- and red-shifted absorption features.} 

\item{Low velocity absorption (LVA) profiles are characterised by an overlapping emission feature and absorption feature, both centred near $\sim$ 0~\kms. These profiles are distinguished from a P-Cygni or Inverse P-Cygni profile due to the presence of emission on both sides of the absoprtion feature.}

\item{Double absorption (DA) profiles are characterised by the presence of two sub-continuum absorption features generally within $\pm$ 200~\kms. There is no emission observed in these profiles.}

\item{Miscellaneous profile types are also observed in the data and are those which cannot be accurately described by any of the above profile types.}

\end{enumerate}

\begin{figure*}[ht]
    \centering
    \includegraphics[width=0.7\textwidth]{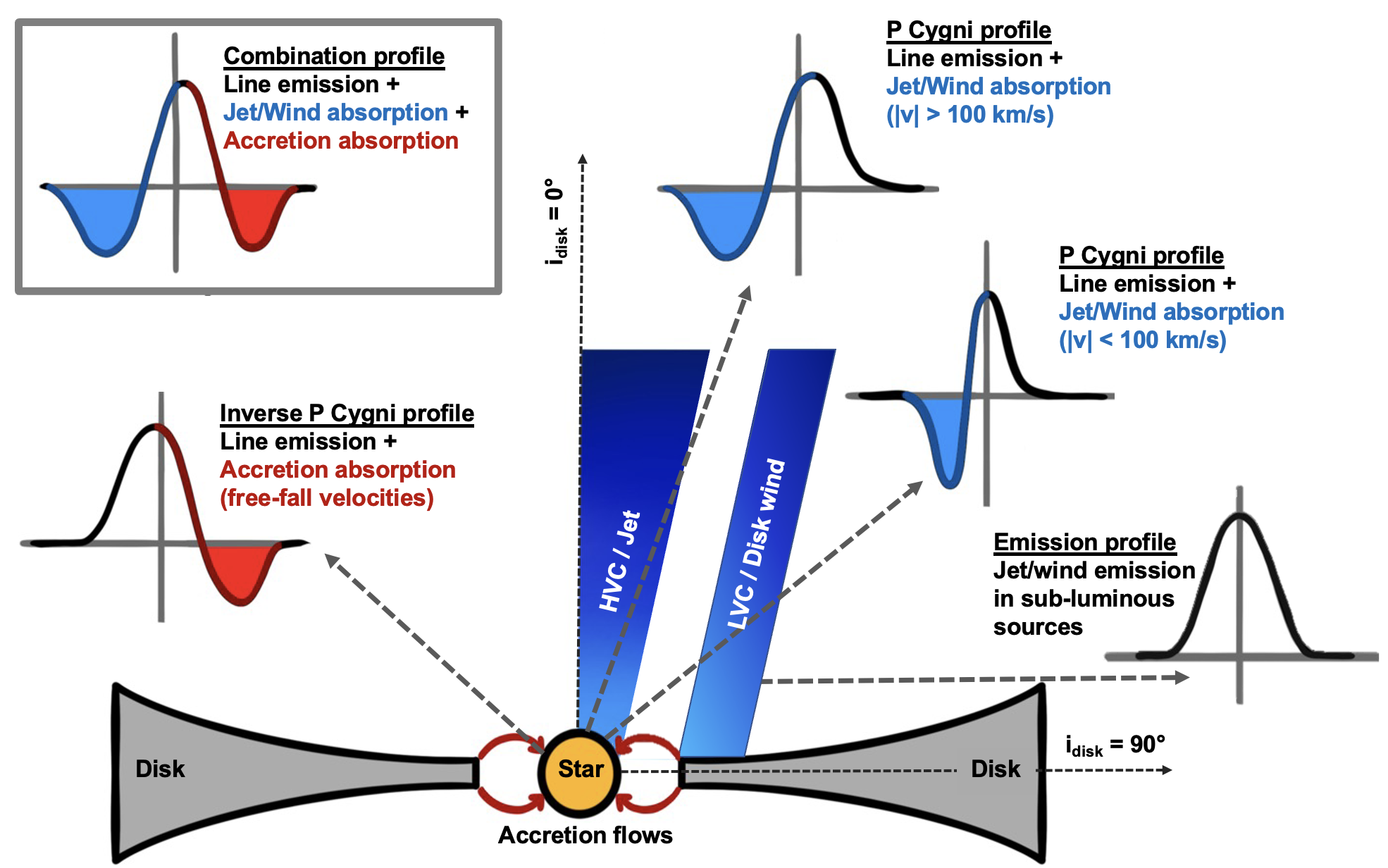}
    \caption{Schematic diagram of various \henir line profiles observed along different line of sights (grey dashed arrows) depending on source inclination (based on \citet{McJunkin14,Xu2021}).}
    \label{fig:sketch}
\end{figure*}

Figure \ref{fig:sketch} shows a sketch of an example protostellar system with accretion flows along which material moves from the inner disk onto the star, and with a jet and disk wind present. The dark grey dashed arrows show different viewing angles which result in different \henir line profiles. In the scenario where the source is viewed along a line-of-sight passing through an accretion flow, a red-shifted sub-continuum absorption feature is observed in the line profile. 
Emission-only profiles may be observed if the source is sub-luminous and/or close to edge-on resulting in the observation of only jet/wind emission. Close to face-on, the emission is absorbed by the higher velocity components of a jet/outflow which results in wider blue-shifted absorption profiles at larger velocities. Meanwhile, for closer to edge-on observations, the observer's line-of-sight passes through the low velocity parts of the outflow (possibly a disk wind), thus producing a narrower absorption feature at less blue-shifted velocities. In some cases, the line-of-sight may pass through both a jet or wind, and an accretion flow resulting in two absorption features in the line profile (see inset in Figure \ref{fig:sketch}). In this figure, we do not include absorption-only profiles, double absorption profiles or low-velocity absorption profiles since they occur less frequently in our sample. 

\begin{figure*}[ht]
    \centering
    \includegraphics[width=0.3\textwidth]{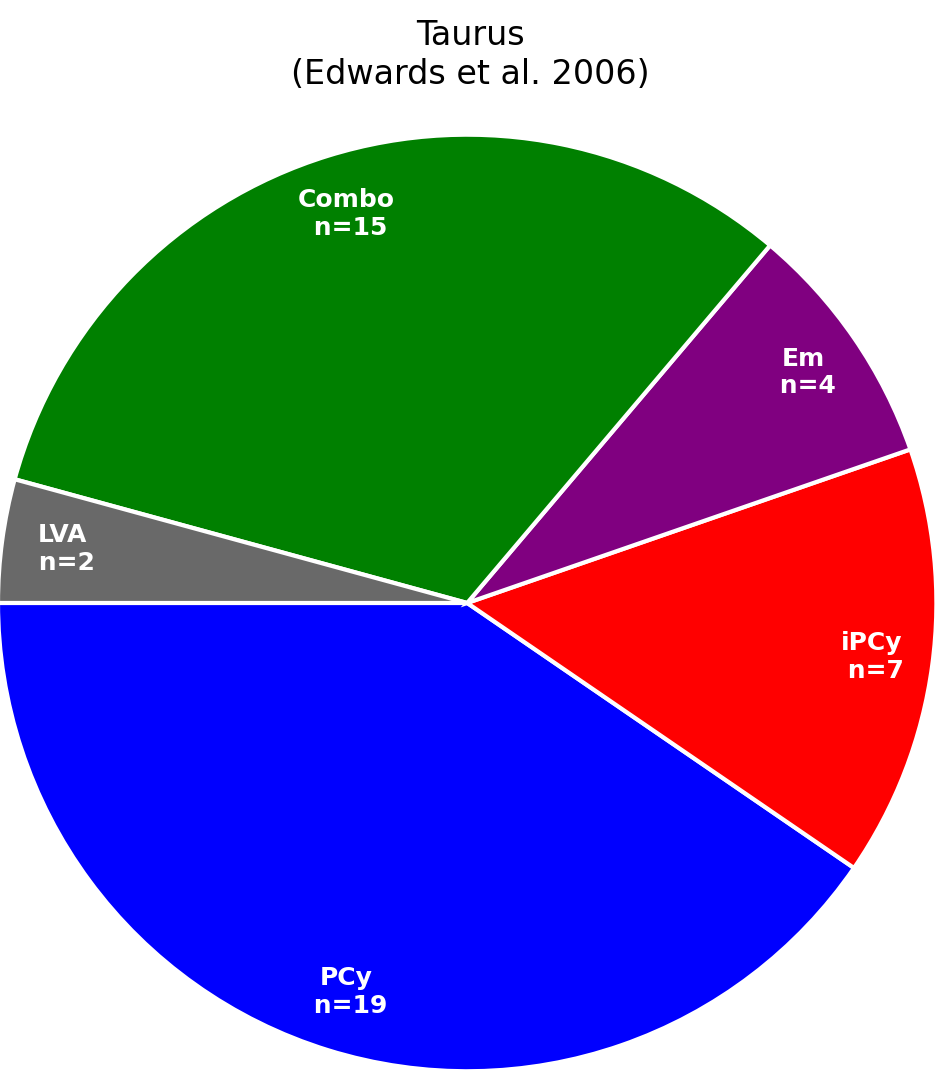}
    \includegraphics[width=0.3\textwidth]{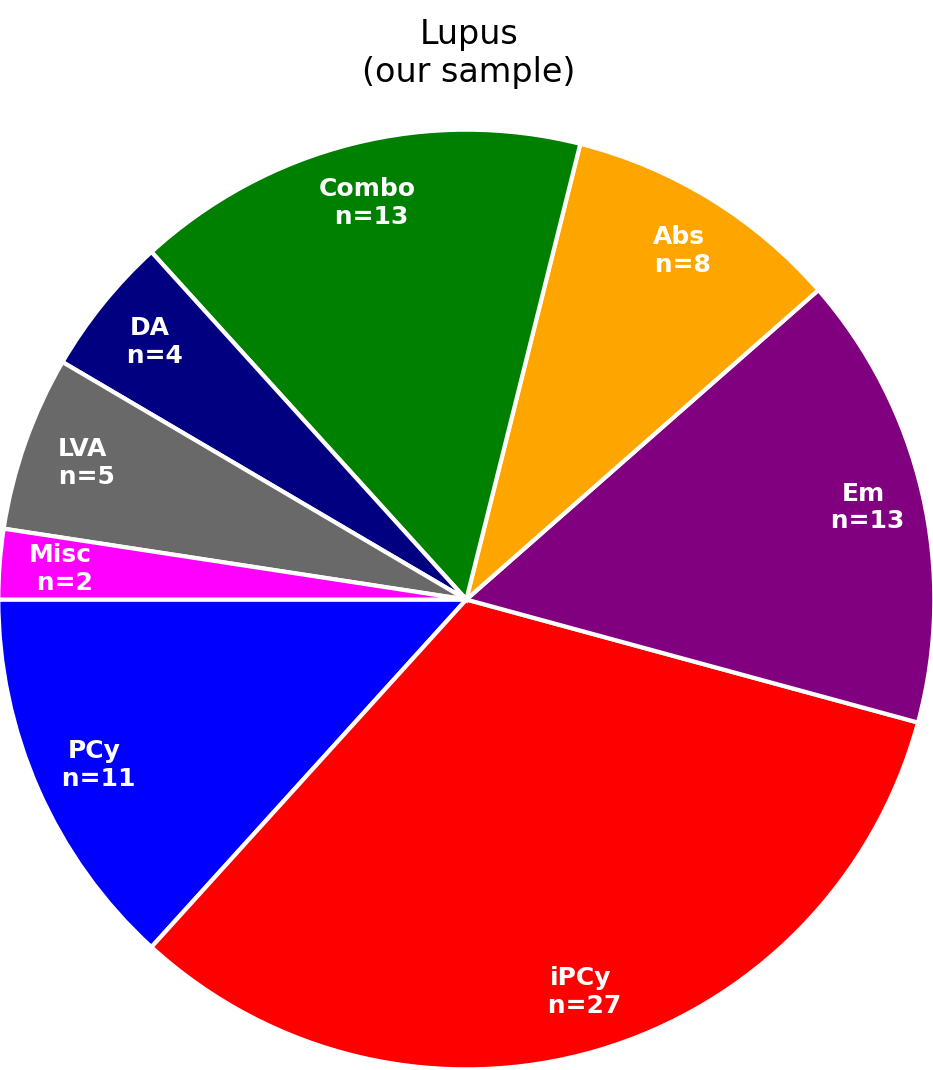}
    \includegraphics[width=0.3\textwidth]{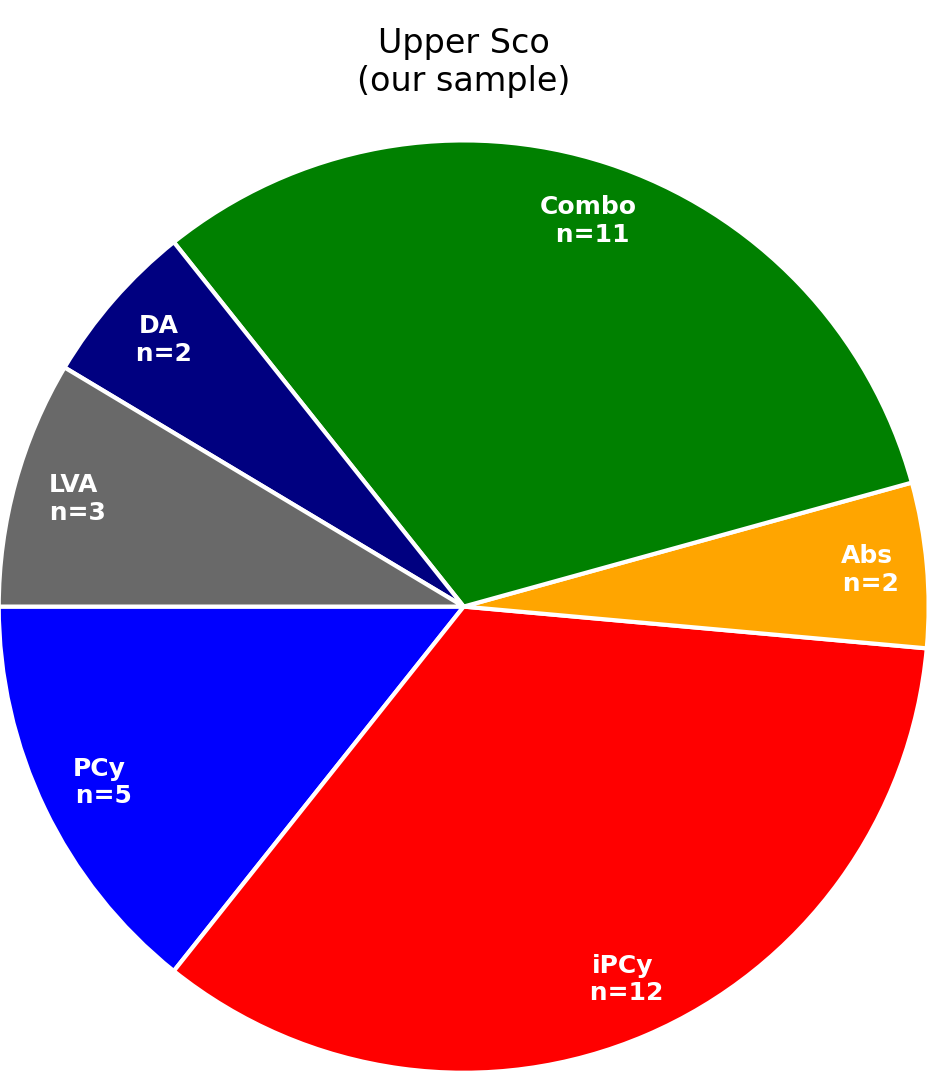}
    \caption{Pie charts showing the proportions of each profile type identified in our sample of 82 sources in Lupus and 35 sources in Upper Scorpius, middle and right respectively. We compare the proportion of profile types observed in our sample to those of a sample of 38 targets located in the Taurus star forming region from \citet{edwards06}. The colours of the wedges in each pie chart represent each profile type - blue P Cygni profiles; red inverse P Cygni profiles; purple pure emission profiles; orange absorption-only profiles; green combination profiles; navy blue double absorption profiles; grey low velocity absorption profiles and pink miscellaneous profiles.}
    \label{fig:piecharts}
\end{figure*}

\section{Results}
\label{sect::Res}

\subsection{Profile appearance statistics}

We fitted the \henir profiles in the Lupus and Upper Scorpius samples as described in Section \ref{sect:fitting}. Figure \ref{fig:piecharts} shows the percentage of targets characterised into each profile type for both regions. Information on the profile type and centroid velocities of the emission and absorption features are listed in Tables \ref{tab:lup_results1} - \ref{tab:usco_results1}. We also compare the proportion of profile types observed in our sample to those of a sample of 38 targets located in the Taurus star forming region from \citet{edwards06}.
In Lupus (middle, Figure \ref{fig:piecharts}) we find that the most common profile type are inverse P Cygni profiles (27 of the 82 targets, i.e. 33\%), followed by combination type profiles (13 targets or 16\%). However in Upper Scorpius (right, Figure \ref{fig:piecharts}) the combination profiles (11 of the 35 targets, 31\%) and inverse P Cygni profiles are the most common types (12 targets, 34\%), followed by P Cygni profiles (five targets, 14\%). In Upper Scorpius we do not observe any pure emission profiles. 
In Taurus, a large portion of the \henir profiles show a P Cygni profile (19 of 38 targets, 50\%), followed by the combination profile type (15 targets, 39\%). There are no single absorption or double absorption profiles observed in the Taurus sample.
We stress that the Taurus sample in \citet{edwards06} (38 objects in total) is much smaller than our sample in Lupus (82 objects), but is comparable in size to our sample in Upper Sco (35 objects). The Taurus sample however is predominantly K7 and M0 type stars, compared to a larger range of spectral types from K0 to M8.5 in our sample. We believe these differences may contribute to the varying proportions of each profile type observed in each region.
Further, we note that the resolution of the Taurus observations is higher than the resolution of our observations. The resolution with NIRSPEC used for the Taurus sample is R $\simeq$ 25000 \citep{edwards06}, while the X-Shooter resolution is $\simeq$ 10000. In velocity, this is roughly equal to 12~\kms and 30~\kms for NIRSPEC and X-Shooter respectively. However, the widths of the features in our sample are typically larger than this resolution, thus we do not believe these variations are due to an instrumental effect. 

\subsection{Absorption velocity properties}
\begin{figure*}[h]
    \centering
    \includegraphics[width=0.7\textwidth]{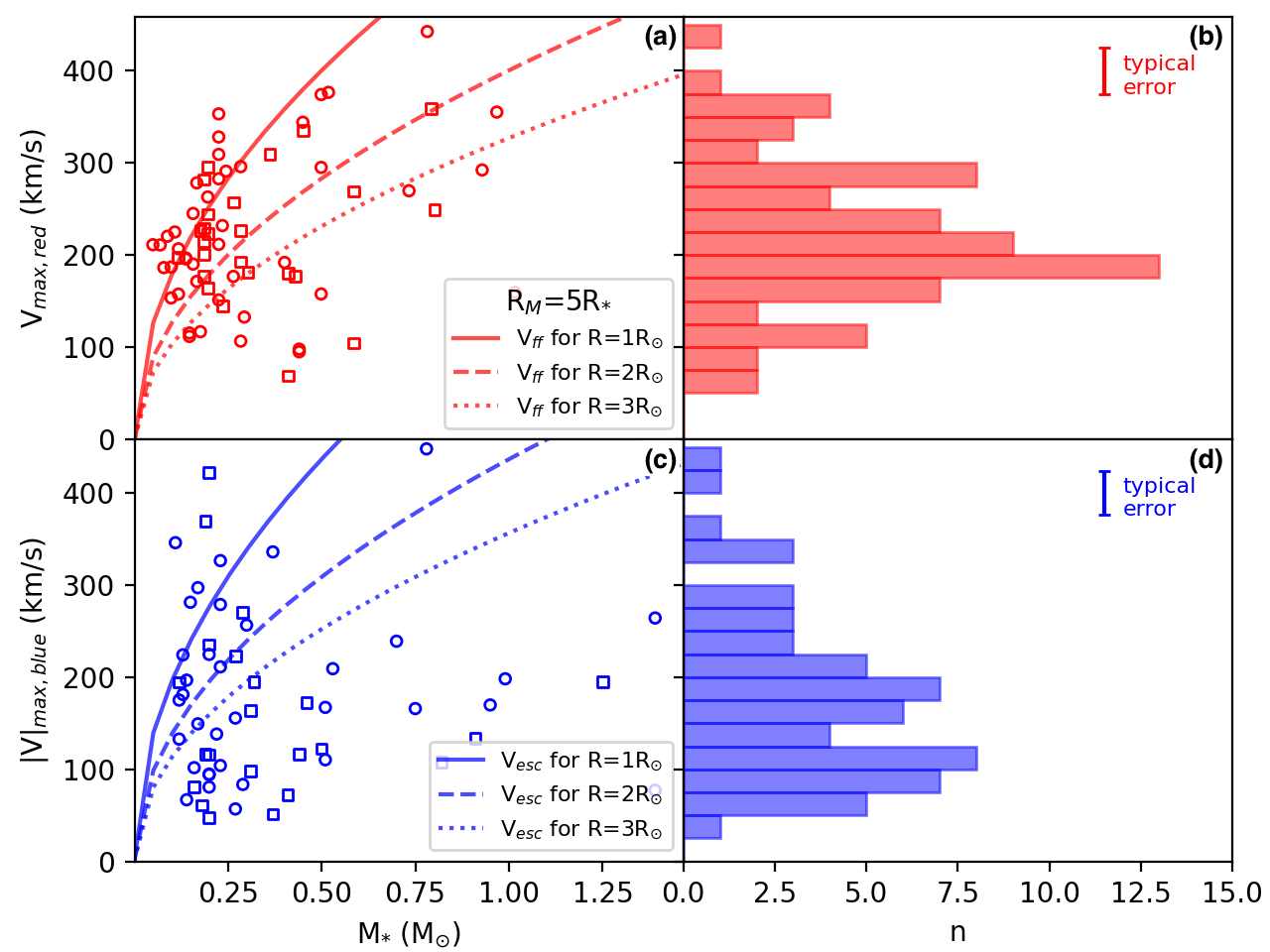}
    \caption{Maximum velocities of the red- and blue-shifted absorption features (panels (a) and (c), respectively) observed in the \henir profiles. Panels (a) and (c) show scatter plots of the Lupus absorption velocities with circles, and Upper Scorpius velocities with squares. The solid, dashed and dotted lines are the free-fall velocities (panel (a)) and escape velocities (panel (c)) for stellar radii R$_{*}$ = 1,~2,~3~R$_{\odot}$, respectively. Panels (b) and (d) show histograms of the maximum absorption velocities for both regions in 10~\kms velocity bins.}
    \label{fig:vels_scat_hist}
\end{figure*}

In Figure \ref{fig:vels_scat_hist} we present scatter plots and histograms of the maximum absorption velocities for the red- and blue-shifted absorption features. The maximum absorption velocity is the highest velocity observed in the absorption feature, i.e. the most blue-shifted velocity in absorption for blue absorption features and the most red-shifted velocity in red absorption features. We calculated the maximum velocities using the fitted Gaussian velocity centroids ($\mu$ or v$_{cen}$) and standard deviations ($\sigma$) i.e. v$_{max}$ = v$_{cen}$ $\pm$ 3$\sigma$. We display the typical 3$\sigma$ fitting errors for the red- and blue-shifted maximum velocities in panels (b) and (d), respectively, which are obtained from the covariance matrix of the Gaussian fit from \textit{scipy curve$\_$fit}.

The majority of red-shifted velocities appear to cluster at velocities of $\simeq$200~\kms, and increase with stellar mass such that they are compatible with free-fall velocities for stellar radii between R$_\star$ = 1 and 2~R$_{\odot}$, assuming the radius for magnetospheric accretion is equal to 5~R$_\star$ \citep{hartmann16}. 
The free-fall velocities are calculated with the following simplified equation from \citet{hartmann16}:
\begin{equation}
\centering
v_{\rm ff} = \left( \frac{2GM_{\star}}{R_{\star}} \right)^{1/2} \left( 1-\frac{R_{\star}}{R_{M}} \right)^{1/2} \simeq 280~M^{1/2}_{0.5}~R_{2}^{-1/2} {\rm km~s}^{-1},
\end{equation}

where G is the universal gravitational constant, R$_{M}$ is the magnetospheric radius from which material falls inwards along accretion flows, and M$_{0.5}$ and R$_{2}$ are the stellar mass and radius in units of 0.5~M$_{\odot}$ and 2~R$_{\odot}$. 
The stellar radii for each source in our sample were calculated using the stellar luminosity and effective temperatures found in \citet{alcala19} for the Lupus sample and \citet{manara20} for the Upper Scorpius sample. The typical R$_\star$ for our targets is $\sim$1.3~R$_{\odot}$ for Lupus and $\sim$1.05~R$_{\odot}$ for Upper Scorpius. We calculated the radius at which the infalling material reaches free-fall velocities (R$_{\rm ff}$) by rearranging Equation 1 in \citet{fischer2008} so that:
\begin{equation}
\centering
    v_{\rm ff} = v_{\rm esc}\left( \frac{R_{\star}}{R_{\rm ff}} - \frac{R_{\star}}{R_{\rm M}} \right)^{1/2}
\end{equation}

becomes

\begin{equation}
    \centering
    R_{\rm ff} = R_{\star}v_{\rm esc} \left( \frac{R_{\rm M}v_{\rm esc}}{1+R_{\rm M}v_{\rm ff}^{2}} \right)
\end{equation}

where R$_{M}$ is set to 5~R$_\star$ following the assumptions of \citet{hartmann16} (see above) and where v$_{esc}$ is the escape velocity defined as:

\begin{equation}
\centering
     v_{\rm esc}~=~\sqrt{2GM_\star/R_\star}
\end{equation}

\begin{figure}[h]
    \centering
    \includegraphics[width=0.39\textwidth]{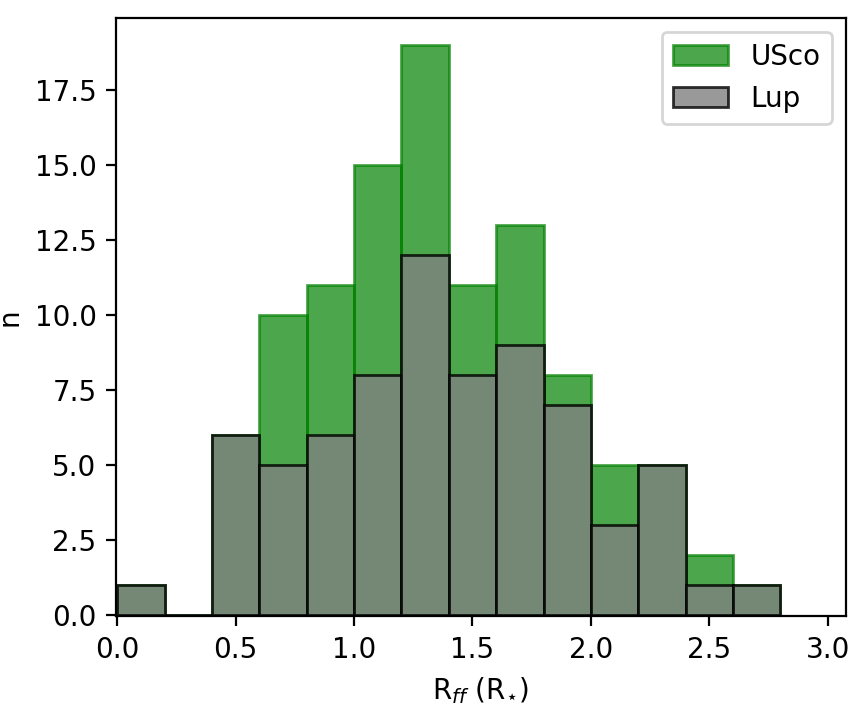}
    \caption{Stacked histogram of the radius in the disk at which the material reaches free-fall velocities. The total value of each bin accounts for sources in both Lupus and Upper Scorpius. }
    \label{fig:rff_hist}
\end{figure}

\begin{figure}[h]
    \centering
    \includegraphics[width=0.35\textwidth]{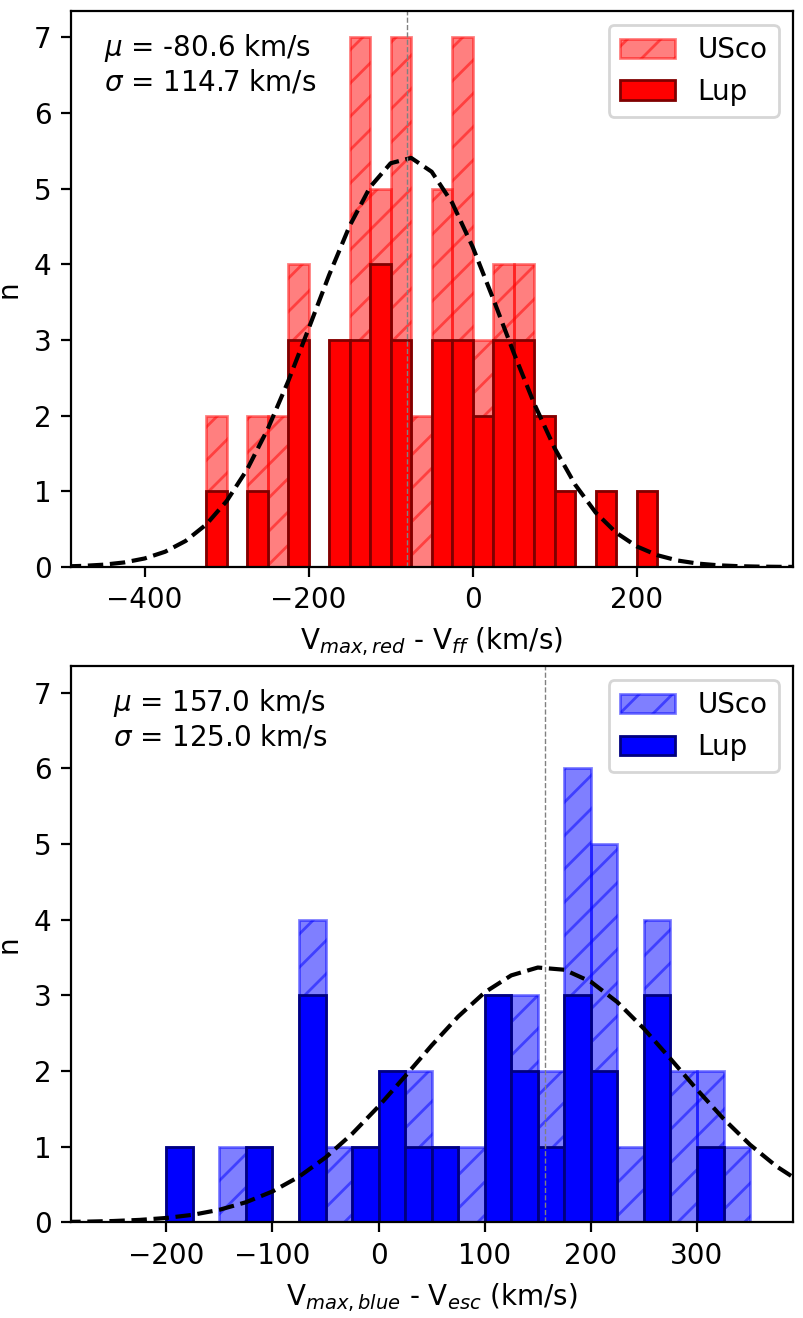}
    \caption{Stacked histograms of the observed red-shifted maximum velocities compared to the free-fall velocities (top), and blue-shifted maximum velocities compared to the escape velocity (bottom), where the total value in each bin is the number of sources in both Lupus and Upper Scorpius. The dashed black line in each panel shows the Gaussian distributions fitted to the histogram data, with the centroid ($\mu$) and standard deviation ($\sigma$) of this fit noted in the top left corner.}
    \label{fig:vmax_vffesc_hist}
\end{figure} 
For sources in Lupus, R$_{\rm ff}$ is approximately 1.6~R$_\star$ while in Upper Scorpius R$_{\rm ff}$ $\sim$ 1.3~R$_\star$ (see Figure \ref{fig:rff_hist}). This typically corresponds to $\sim$ 1.4--2~R$_{\odot}$ in line with the curves in Figure \ref{fig:vels_scat_hist} (see Section \ref{sect:abs_features} for further discussion).

The blue-shifted velocities in our sample tend to appear at lower velocities than their red-shifted counterparts, but are not as confined to specific velocities. Further, they generally reach escape velocities from the star. We note that the escape velocities calculated for our sample are from the star, and that if the material is launched in a disk wind, the escape velocities from the disk are smaller.

For each source in Lupus and Upper Scorpius, in Figure \ref{fig:vmax_vffesc_hist}, we plot histograms of the difference between the observed maximum red-shifted velocities and free-fall velocity, and between the maximum blue-shifted velocities and escape velocity from the star. The red-shifted maximum velocities appear to be similar to the free-fall velocities, with a mean value of -80.6~\kms.
Meanwhile, the blue-shifted velocities are generally larger than the escape velocity with many of the values of v$_{\rm max}$-v$_{\rm esc}$ between 100-200~\kms (and a mean value of 157~\kms), but their distribution is flatter spanning a large value of mostly positive velocity differences.

\begin{figure}[h]
\centering
\includegraphics[width=0.39\textwidth]{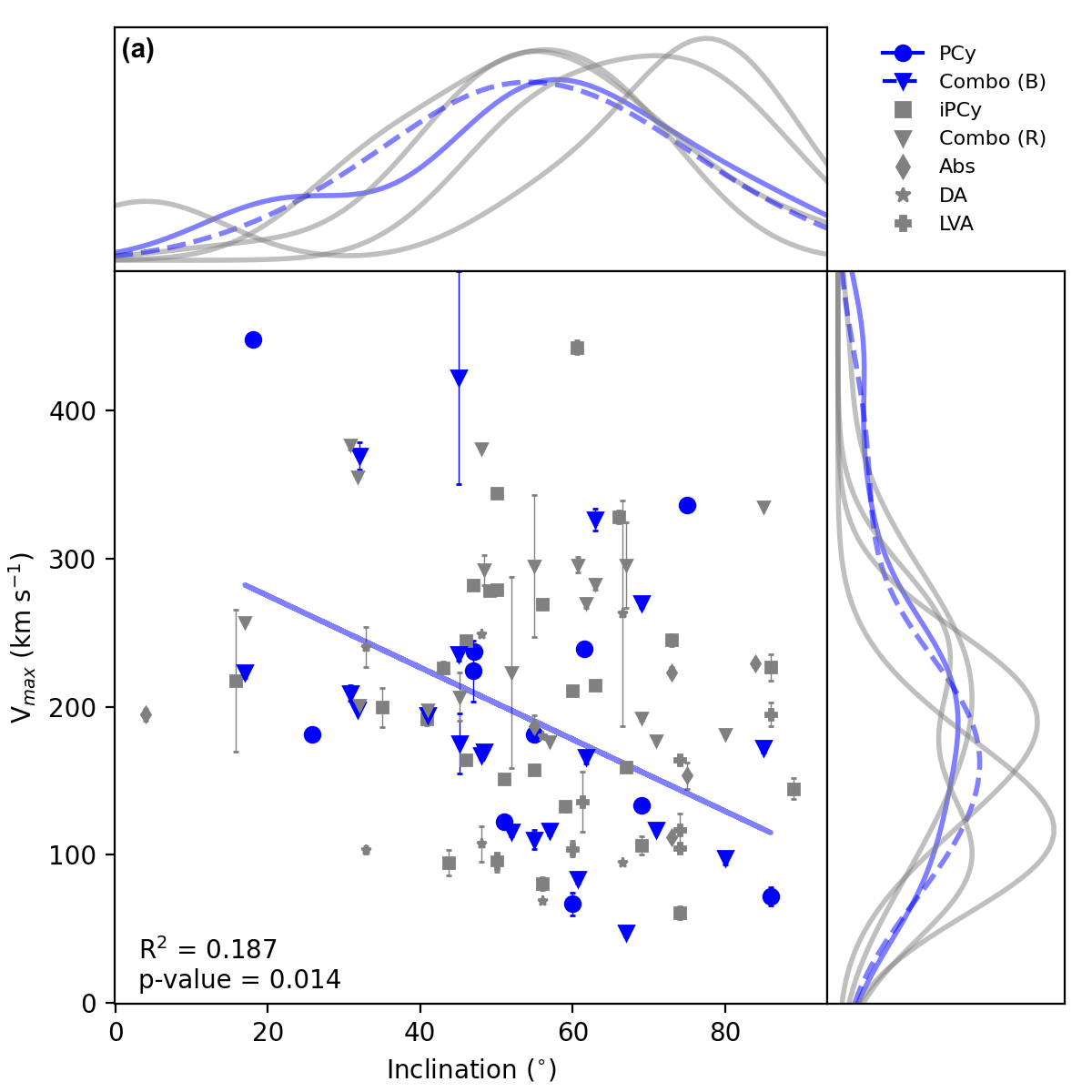}\\
\includegraphics[width=0.39\textwidth]{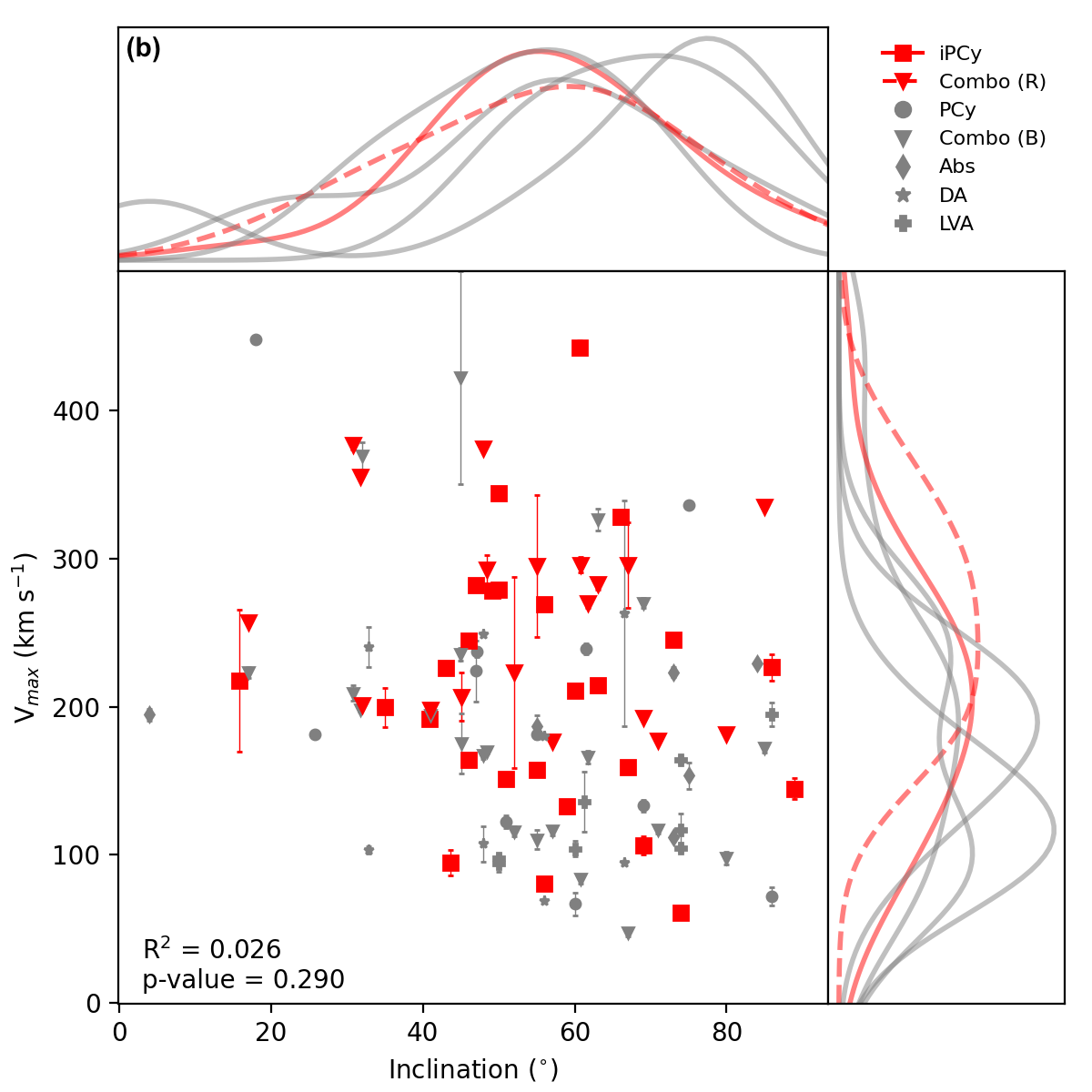}
\caption{Source inclinations with respect to the maximum velocities of observed absorption features (a) for the blue absorption features in P Cygni and combination profiles, and (b) for the red absorption features in inverse P Cygni and combination profiles in our sample. All other profile types are displayed in grey. 
The curves along the top and right-hand side are the kernel density estimation (KDE) of each profile type.}
\label{fig:fit_vmax_inc}
\end{figure}

\begin{figure}[h]
\centering
\includegraphics[width=0.39\textwidth]{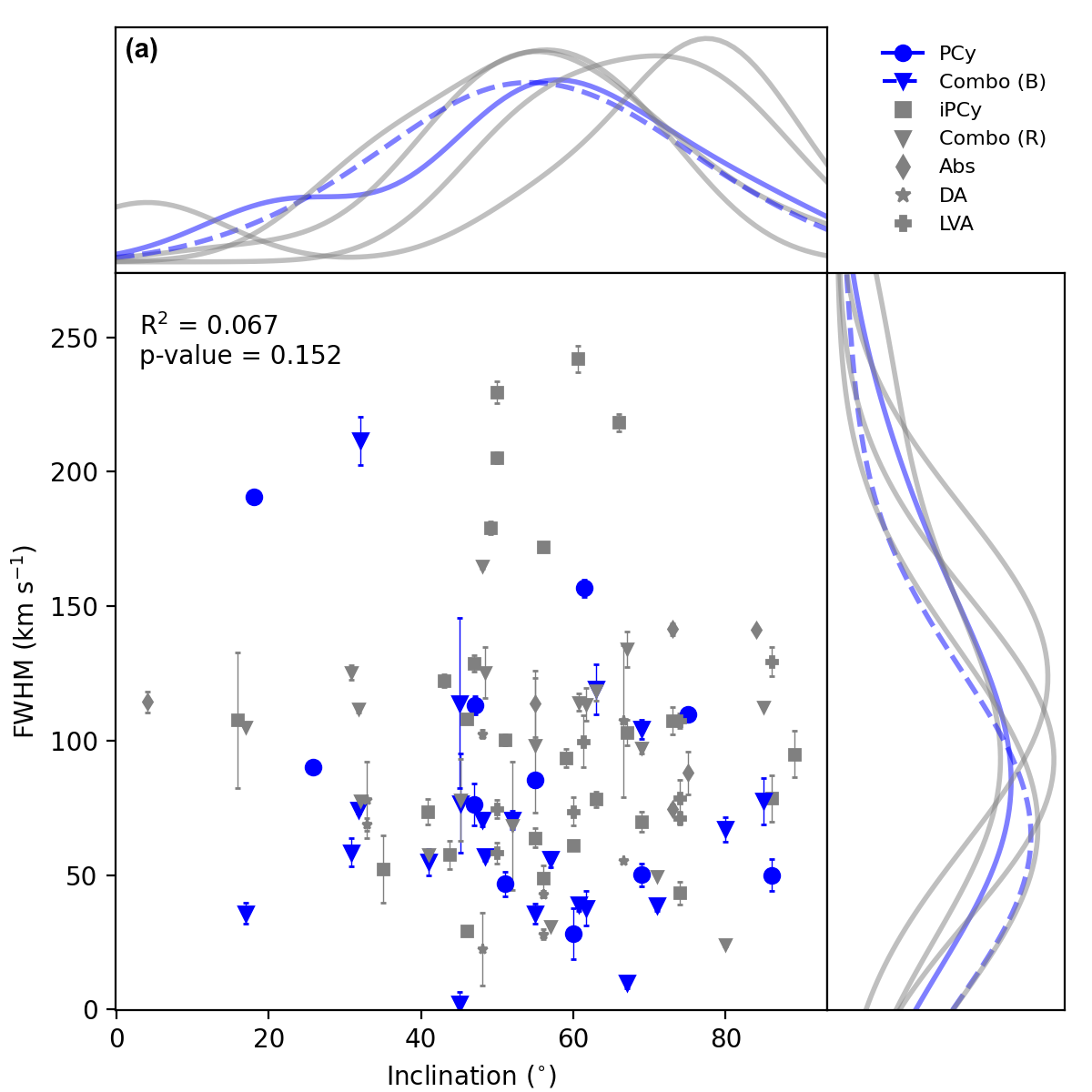}\\
\includegraphics[width=0.39\textwidth]{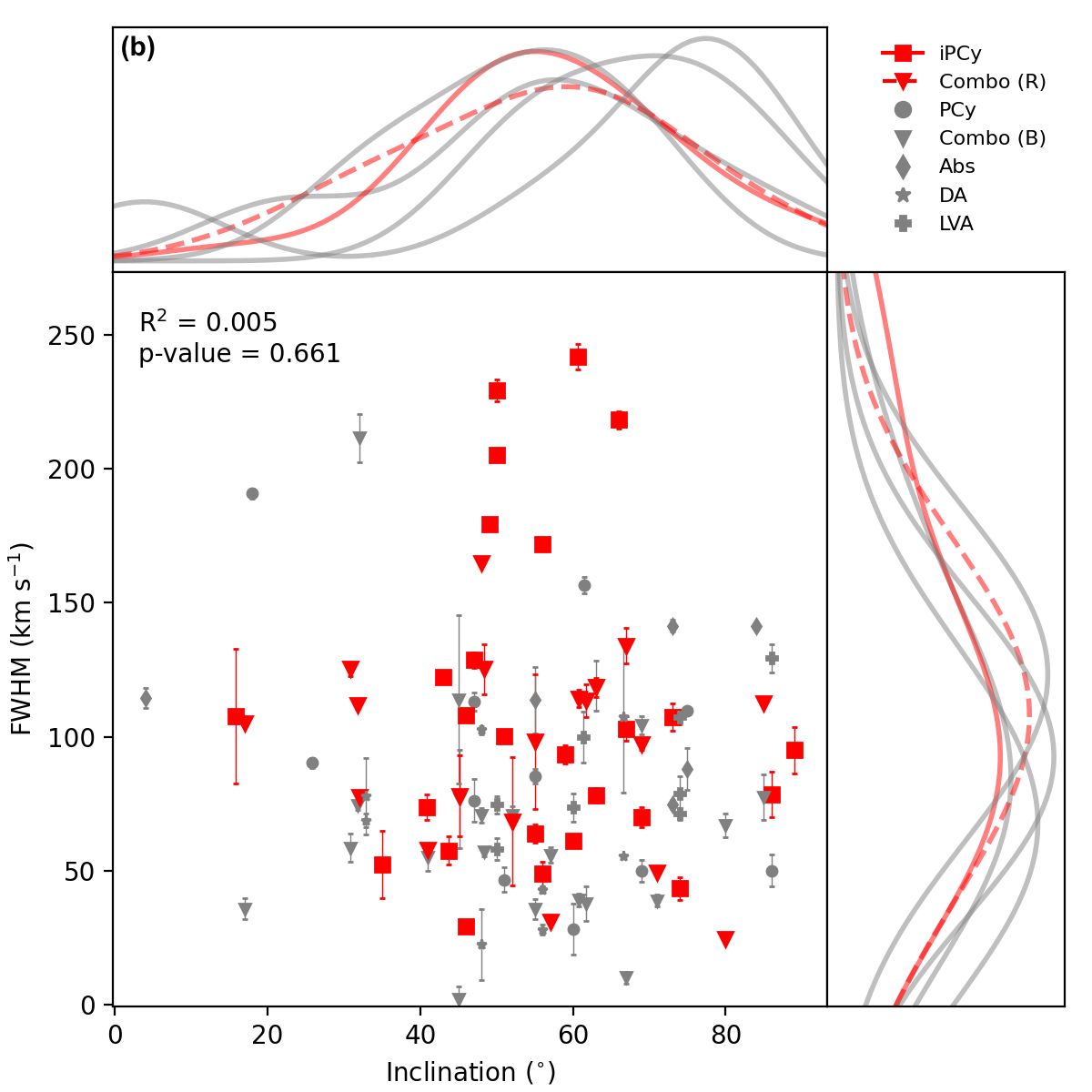}
\caption{Source inclinations with respect to the full-width half maximum of observed absorption features (a) for the blue absorption features in P Cygni and combination profiles, and (b) for the red absorption features in inverse P Cygni and combination profiles in our sample. All other profile types are displayed in grey. The curves along the top and right-hand side are the kernel density estimation (KDE) of each profile type.}
\label{fig:fwhm_vmax_inc}
\end{figure}

In Figures \ref{fig:fit_vmax_inc} and \ref{fig:fwhm_vmax_inc} we present the maximum absorption velocities and the FWHM of the absorption features with respect to the inclination of the source for both the Lupus and Upper Scorpius regions. 
In Figure \ref{fig:fit_vmax_inc}, we observe the maximum blue-shifted velocity trending to lower velocities at higher inclinations (panel (a)). This is not seen in their red-shifted counterparts, as the red-shifted velocities remain $\geqslant$ 150~\kms\ (panel (b)), even for highly inclined targets. 
In combination profiles (downward triangles), most blue absorption features appear to have a maximum velocity of $\simeq$ 150 - 200~\kms, while a second cluster of velocities closer to 350~\kms\, which are generally the red-shifted absorption feature for these profiles. Given that the velocities for combination profiles appear across all inclinations, we believe that they are not due only to a particular viewing angle. Red-shifted absorption features tend to occur at higher velocities than the blue absorption features, and they also appear generally wider (see Figure \ref{fig:fwhm_vmax_inc}). This may suggest that we can observe a wider range of accretion velocities than velocities in the wind, hinting at a dependence on the source inclination. Red-shifted velocities, however, are not related to the source inclination, like the blue-shifted features as evident in the p-values displayed in each panel. In Figure \ref{fig:fwhm_vmax_inc} we find that the red-shifted absorption features are wider (typical FWHM $\simeq$ 100~\kms) than the blue-shifted features (typical FWHM $\simeq$ 50--100~\kms). This is noticeable even in the combination profiles (downwards triangles). However, we do not find a correlation between the width of the absorption features and the source inclination. 

\section{Discussion}
\label{sect::disc}


In the following sections we discuss the results presented in Section \ref{sect::Res}. We begin by focusing on the \henir profiles with pure emission profiles and the presence of known jets in these sources. Following this, we examine the profiles with absorption features for trends related to the source inclination and accretion properties. Finally, we examine the evolutionary trends observed in the data.

\subsection{Pure emission} \label{sect:emonly}

We observe 12 sources in Lupus without sub-continuum absorption features. Emission in the \henir line arises due to scattering and in situ emission. Pure emission profiles may arise due to a stellar wind, launched from high latitudes on the star and observed at typical inclinations of $\simeq$ 70$^{\circ}$ \citep{kwan2007}. 
Of the 12 stars with emission-only profiles, six targets have known jets. These are: Par-Lup3-4 \citep{whelan2014}; Sz102 \citep{murphy2021}; Sz69, Sz73, Sz123A and Sz123B which show both a high-velocity component (HVC) and low velocity component (LVC) in the [OI]$\lambda$6300\AA\, line (\citealt{nisini18}). The HVC is generally attributed to high-velocity jets which can produce either emission-only profiles (for edge-on sources) \textit{or} P Cygni profiles (closer to pole-on) - see Section \ref{sect:abs_features}. A further four sources show signatures of a LVC, possibly a disk wind - these sources are Sz88A, Sz106, Sz133 and RX1556.1-3655 \citep{nisini18}. 

Five sources (Par-Lup3-4, Sz102, Sz106, Sz123B, Sz133) are sub-luminous (sl) sources, which have a stellar luminosity much lower (by at least a factor ten) than other sources of the same spectral type \citep{alcala14, nisini18}. 
Only three of the sub-luminous sources (Sz102, Sz123B and Sz133) have known inclinations, all of which are greater than 40$^{\circ}$.
Sub-luminous sources are generally expected to be observed edge-on \citep{hughes94} but it is possible that the disk inclinations reported for these sources are underestimated, or that there is a misalignment between the inner and outer disk in some of these sources \citep{bohn2021} as previously mentioned. Jet morphology and proper motions of the Sz102 jet, for example, indicate that the jet is in the plane of the sky \citep{murphy2021}, i.e. the disk should be edge-on, thus revealing a possible misalignment between the inner and outer disk. 
Additionally, there are two other sub-luminous sources in the Lupus sample - these are Lup706 (an inverse P Cygni profile) and SSTc2dJ160703.9-391112 (absorption-only profile), so the lower luminosity is not unique to sources showing the emission-only profiles.
Three of the sources with pure emission profiles exist in binary systems - these are Sz88A, and Sz123A and Sz123B \citep{alcala14}. The inclination of the remaining source (J16085953-385627) with a pure emission profile is not known and it doesn't appear to be associated with any known jet. 

\begin{figure}[h]
    \centering
    \includegraphics[width=0.35\textwidth]{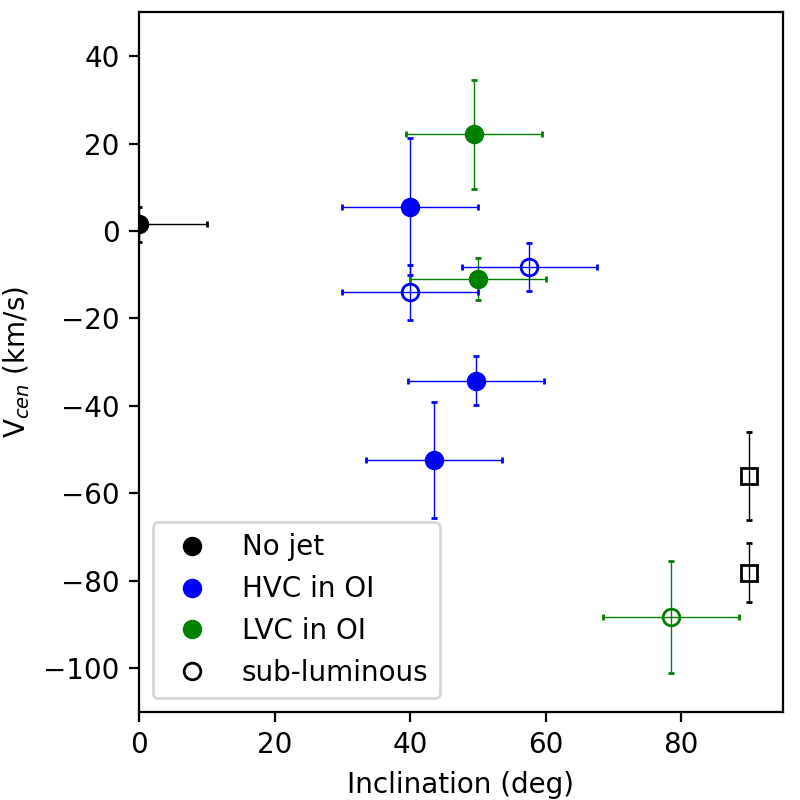}
    \caption{Centroid velocities of the emission-only profiles versus the source inclination. Sub-luminous sources are plotted with the hollow markers. Square markers indicate sub-luminous sources with unknown inclinations.}
    \label{fig:em_vcen_inc}
\end{figure}

\begin{figure}[h]
    \centering
    \includegraphics[width=0.35\textwidth]{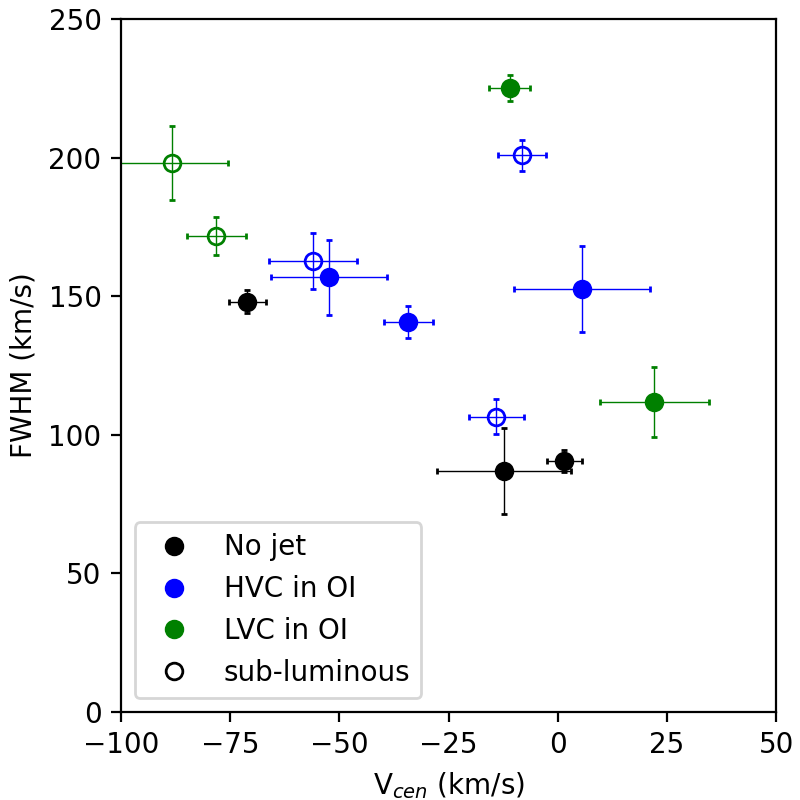}
    \caption{FWHM versus the centroid velocity of the emission-only profiles. Sub-luminous sources are plotted with hollow markers.}
    \label{fig:em_vcen_fwhm}
\end{figure}

We searched for a correlation between the centroid velocity of the emission profiles and the source inclination (see Figure \ref{fig:em_vcen_inc}). Generally, the centroid velocities were blue-shifted by up to 100~\kms, and those at higher velocities are narrower than the emission profiles closer to 0~\kms (see Figure \ref{fig:em_vcen_fwhm}). The fact that we tend to observe emission-only profiles for sources with known jets and at high inclinations lends support to the models of \citet{kwan2007}. They note that when viewing the \henir line for sources at higher inclinations, a stellar wind, which appears as a P Cygni profile for lower inclinations, may become an emission-only profile as the blue-shifted absorption feature becomes weaker and eventually disappears (see Figure \ref{fig:sketch}). However, we note that for the 70 remaining sources in Lupus, 48 of the sources (68\%) have a LVC and/or HVC signature in [OI] \citep{nisini18}. HVC signatures in [OI] are observed in sources with P Cygni profiles in the \henir line at similar rates as the emission-only profiles - seven out of 11 P Cygni profiles in Lupus were observed with a HVC in [OI], compared to six of 12 emission-only profiles. Only ten sources with other profile types, however, are observed with a HVC in [OI].
Of the sources plotted in Figure \ref{fig:em_vcen_inc}, only one is not observed with either a LVC or HVC in OI - this source is Lup607 which has an inclination of 0$^{\circ}$. The inclinations of the other two sources without a known jet or outflow (J16085953-3856275 and SSTc2dJ160708.6-391408) are not known. Further, in Figure \ref{fig:em_vcen_inc}, we plot the sub-luminous sources without known inclinations at 90$^{\circ}$ (in line with the assumption that sub-luminous sources are often viewed edge-on). These sources show centroid velocities that are blue-shifted to velocities up to -80~\kms, in line with the apparent correlation of centroid velocity and source inclination for the emission profiles. However, if the inclinations for the two sub-luminous sources located at $\sim$~40-60$^{\circ}$ are underestimated, 
the correlation between centroid velocity and inclination is less evident if these sources are closer to edge-on ($>$ 70$^{\circ}$).
It is also possible that disk occultation may result in the emission profile becoming blue-shifted, which could explain a trend in the centroid velocity with respect to source inclination. 

In a number of emission-only sources (e.g. Sz73, Sz88A, Sz102 and SSTc2dJ160708.6-391408), we observe asymmetric line profiles as is often seen in forbidden emission lines (for example, by \citealt{nisini18,banzatti19}). 
While the majority of our targets with emission-only profiles are associated with known jets, these profiles can still be produced in other scenarios. For example, in Figures \ref{fig:Lup1} and \ref{fig:Lup2}, where we show all Lupus profiles arranged by profile type, we see a number of asymmetric emission profiles (e.g. Par-Lup3-4, Sz73, Sz106, Sz133 and RX1556.1-3655), which appear to have low-velocity absorption features that are not sub-continuum, and thus not fitted by our fitting routine. This is also the case in the Taurus sample studied by \citet{edwards06}, where four targets show asymmetric emission profiles indicating possible red and/or blue absorption that is not sub-continuum. 
Overall, we find that sources showing an emission-only profile tend to be associated with known jets and outflows, and appear at high source inclinations.

\subsection{Profiles with absorption features}
\label{sect:abs_features}

We observe absorption features in many spectra from both regions in our sample. 
In the following sections we discuss observed trends in velocities of the wind and accretion features, and investigate if the measured absorption velocities have any dependence on the source inclination and/or accretion properties.


\subsubsection{Velocity of the absorption features}
\paragraph{\textbf{Absorption features tracing winds}}

In Figure \ref{fig:vmax_vffesc_hist}, we present histograms of the difference between our observed blue-shifted velocities and the escape velocities from the star.
Eleven sources show v$_{max}$ lower than the escape velocity (i.e. below 0~\kms in Figure \ref{fig:vmax_vffesc_hist}), of which seven sources have known inclinations. These sources occur across the range of source inclinations, so this is not thought to be an effect of the viewing angle. Thus where the measured velocities are lower than the escape velocity, we suggest that not only gravitational forces act on the wind, but that there may also be magnetic fields at play. In wind launching models, it is assumed that the wind can be accelerated to a few times the escape velocity at the launching radius, especially in MHD launching models (see review by \citealt{pudritz2007}). However, if we are tracing disk winds, the escape velocity from the disk would be much lower than the stellar escape velocity.
It is possible to distinguish between which type of wind based on the widths of the blue absorption features in the \henir profiles. 
Narrower blue absorptions at low velocities are generally attributed to disk winds, while wider features across a range of velocities are more likely to originate in stellar winds \citep{edwards06,kwan2007}. 

\paragraph{\textbf{Absorption features tracing accretion}}
In Figure \ref{fig:vmax_vffesc_hist}, we also plot the difference between our observed red-shifted velocities and the free-fall velocities. In the top panel, 
a number of sources have observed red-shifted velocities less than free-fall velocities. \citet{hartmann16} note that magnetospheric accretion models assume free-fall along axisymmetric, dipolar magnetic field lines at a constant infall rate. For measured $v_{\rm max}$ lower than the expected free-fall velocities, it is possible that the material has simply not yet reached its terminal speed. \citet{cauley14,cauley15} suggest that the maximum velocities can be lower than free-fall velocities in cases where the magnetospheric accretion geometry is more compact due to a weaker magnetic field. 
On the other hand, for larger values of $v_{\rm max}$, the general assumptions of axisymmetric and dipolar magnetic fields may not apply and more complex fields may be present. For example, \citet{gregory12} have derived magnetic field maps for a number of T Tauri stars and find that the large-scale magnetic field topology can vary substantially with stellar properties, often finding non-axisymmetric fields with strong multipolar components. Further, typical 3$\sigma$ errors on the measured velocities are, on average, a few tens of \kms, which cannot explain why some sources have $v_{\rm max}$ approximately 100~\kms greater than the free-fall velocity.

Thus, the velocities of the absorption features in the \henir line profiles may reveal information about the geometry of both the outflow(s) and the accretion flows in the system.

\subsubsection{Dependence on source inclination}

The inclination of the source can affect the observed profiles, thus in Figures \ref{fig:fit_vmax_inc} \& \ref{fig:fwhm_vmax_inc} we compare the maximum velocity and the widths of the absorption features to the inclination. In these figures, the curves along the top and right-hand side are the kernel density estimation (KDE) of each profile type which we use to estimate, for example, the most likely inclination at which a certain profile type is observed. The blue points in panel (a) represent the P Cygni profiles and blue feature in combination profiles, while the red points (panel (b)) represent the inverse P Cygni profiles and the red feature in combination profiles. Only the blue or red points are fitted with a line of best fit. The grey points in these panels represent all the other profile types, and are plotted only for comparison with the highlighted blue or red data points. We use the R$^{2}$ values and p-values as a measure of the goodness of each fit. 
The R$^{2}$ value is a measure of the scatter in our data, while the p-value measures the likelihood that any observed trends are significant and not a result of chance. A high R$^{2}$ indicates a small scatter and a good linear fit, but we tend to observe low R$^{2}$ values due to large scatter. Instead, the p-value gives us a better indication as to whether or not there is a significant trend between, for example, the maximum velocity and the inclination. Generally, a p-value lower than 0.05 is accepted as evidence of a relationship between two variables.

\paragraph{\textbf{Absorption features tracing winds}}
With the exception of a few sources, blue-shifted absorption features are generally not observed at low inclinations ($<$ 40$^{\circ}$). However, we do see a slight trend in the blue-shifted velocities, as the lowest velocities are present at high inclinations, suggesting that we observe different velocity components of the wind at different relative inclinations to the star. 
This dependence on inclination is also seen in models of the \henir line \citep[e.g.][]{kwan2007}, and is likely due to changing line-of-sight velocities when viewing the wind at different inclinations \citep{kurosawa12}. This is illustrated by \citet{Xu2021} and in Figure \ref{fig:sketch}, where at higher inclinations (edge-on) absorption features are created by lower velocity disk winds leading to less blue-shifted and narrow features. 
The blue-shifted features appear narrower than their red-shifted counterparts, which was generally seen in previous observations of a smaller sample \citep{edwards06}. In Figure \ref{fig:fwhm_vmax_inc} panel (a), a number of blue-shifted absorption features have FWHMs of $\sim$ 50~\kms, particularly for combination profiles. We note however that the blue-shifted features become slightly narrower at high inclinations, thus knowledge of the viewing angle is important to determine which type of wind is present. 
Our measured velocities and widths of the blue-shifted absorption features are consistent with previous studies, for example \citet{kwan2007}, that find that the blue-shifted features in P Cygni profiles are expected to become less blue-shifted and narrower as the inclination changes from face-on (0 degrees) to edge-on (90 degrees).

\paragraph{\textbf{Absorption features tracing accretion}}
Red-shifted velocities appear to be more independent of the source inclination. 
We see a majority of red-shifted velocities near $\simeq$200 -- 250~\kms\, tailing off to higher velocities up to $\simeq$400~\kms, but with no significant decrease in velocity at higher inclinations. Further, these velocities are consistent with free-fall velocities for stellar radii between R$_\star$ = 1 -- 2~R$_{\odot}$ assuming a magnetospheric radius of R$_{M}$ = 5~R$_\star$ \citep{hartmann16} (see Figure \ref{fig:vels_scat_hist}). This is similar to the calculations of \citet{kwan2011} who found that infalling gas can reach speeds of 150~\kms \, near 2~R$_\star$ if the accretion flow starts between 4 -- 8 R$_\star$. 
Previous models of the red-shifted absorption in the \henir line for a star with a dipolar magnetic field showed narrowing of the absorption at higher inclinations \citep{fischer2008}, suggesting that the sources in our sample may have a more complex magnetic field than the dipolar axisymmetric fields normally assumed in models of magnetospheric accretion \citep{hartmann16}. We do not observe such a trend in the measured velocities and widths of the red-shifted absorption features in our sample. We also reiterate that the inclinations presented in these plots are for the outer disk, which may be misaligned from the inner disk \citep{bohn2021}.

\subsubsection{Dependence on accretion properties}

\begin{figure}[h]
\centering
\includegraphics[width=0.4\textwidth]{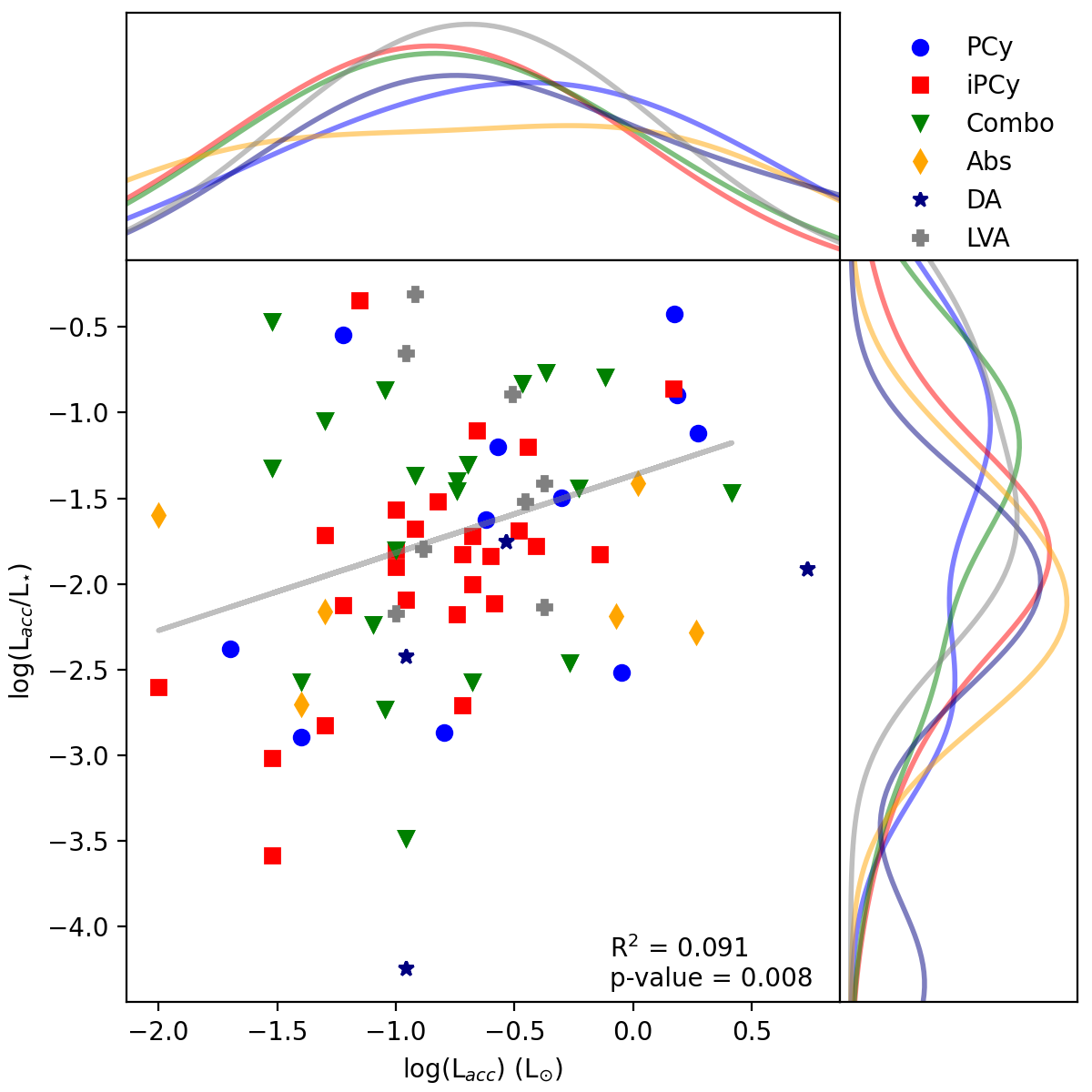}
\caption{L$_{\rm acc}$/L$_{\star}$ versus the accretion luminosity L$_{\rm acc}$ for all profile types with absorption features.}
\label{fig:LMacc_LMstar}
\end{figure}

In Figure \ref{fig:LMacc_LMstar}, for each profile type we plot the ratio of the accretion luminosity to the stellar luminosity with respect to the stellar luminosity.
P Cygni profiles are observed over a range of accretion luminosities, while inverse P Cygni and combination profiles have stronger peaks in their distribution across the range of accretion luminosities. 
The ratio of L$_{\rm acc}$/L$_{\star}$ varies for each of these profile types, with a bimodal distribution in the P Cygni profiles (peaking at log(L$_{\rm acc}$/L$_{\star}$) $\simeq$ -1.0 and a second peak at $\simeq$ -3.0), however low number statistics brings into question the significance of the bi-modality. Single peaks are observed for the inverse P Cygni profiles (at log(L$_{\rm acc}$/L$_{\star}) \simeq$ -2.0) and combination profiles (at log(L$_{\rm acc}$/L$_{\star}) \simeq$ -1.2). These results appear to show that profiles with blue-shifted absorption features, tracing the presence of an outflow, occur in sources with higher accretion rates. 
Recent models of disk evolution have shown that MHD disk winds in particular can account for the observed accretion properties of young stars in Lupus \citep{tabone21}, suggesting that accretion is an important part of driving winds and outflows. If these winds are indeed accretion-powered, as suggested by \citet{edwards06}, this result shows that the \henir line is a useful way to probe the connection between accretion and ejection.

We find that targets with blue absorption features (i.e. P Cygni or combination profiles) have slightly stronger accretion than those targets with only red absorption (i.e. inverse P Cygni profiles). This however seems to be driven mostly by the number of combination profiles present at higher accretion luminosities. Nonetheless, the higher accretion rates observed for targets with blue absorption features supports the findings of \citet{edwards06}, who observe stellar winds at targets with a wide range of accretion rates. 

\begin{figure}[h]
\centering
\includegraphics[width=0.4\textwidth]{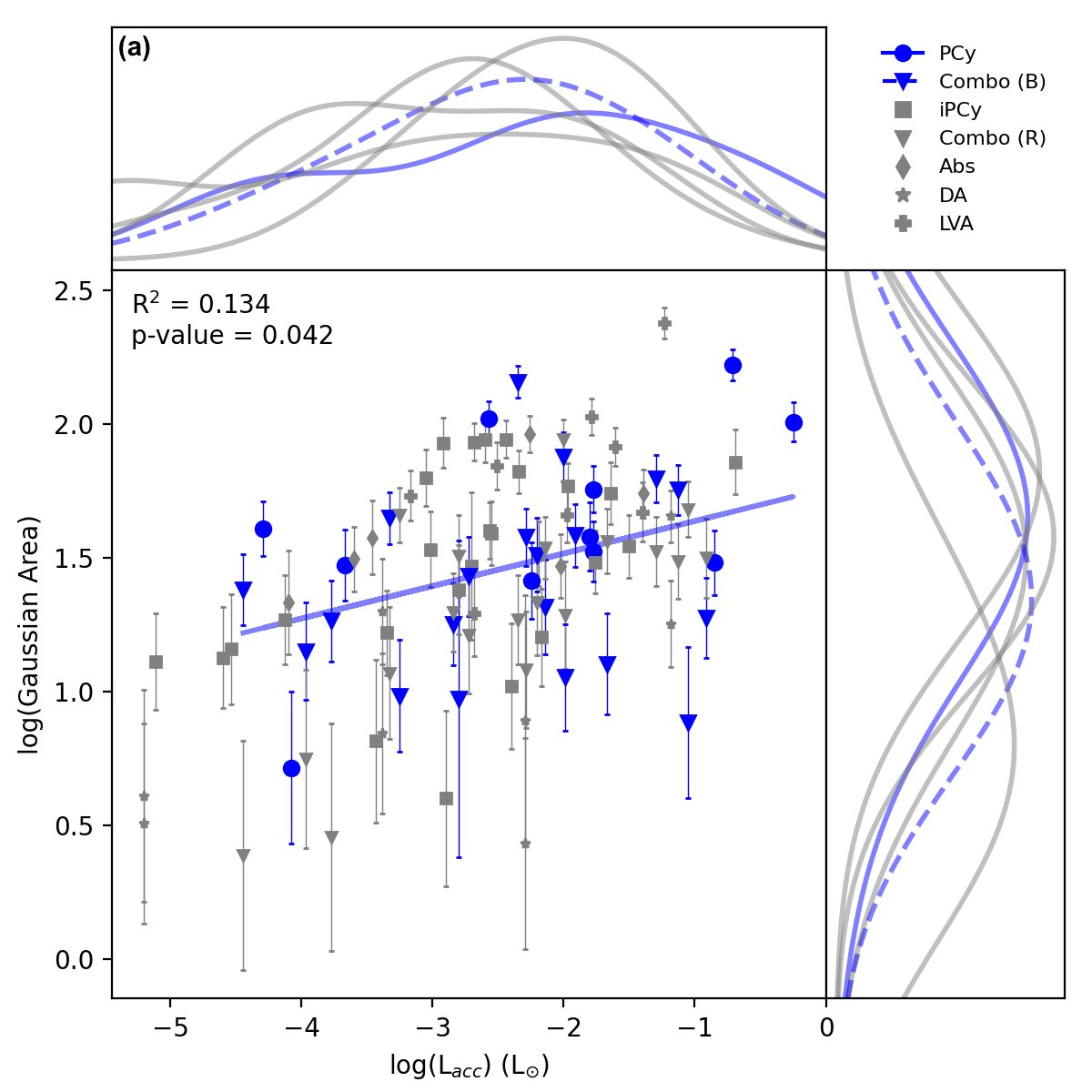}\\
\includegraphics[width=0.4\textwidth]{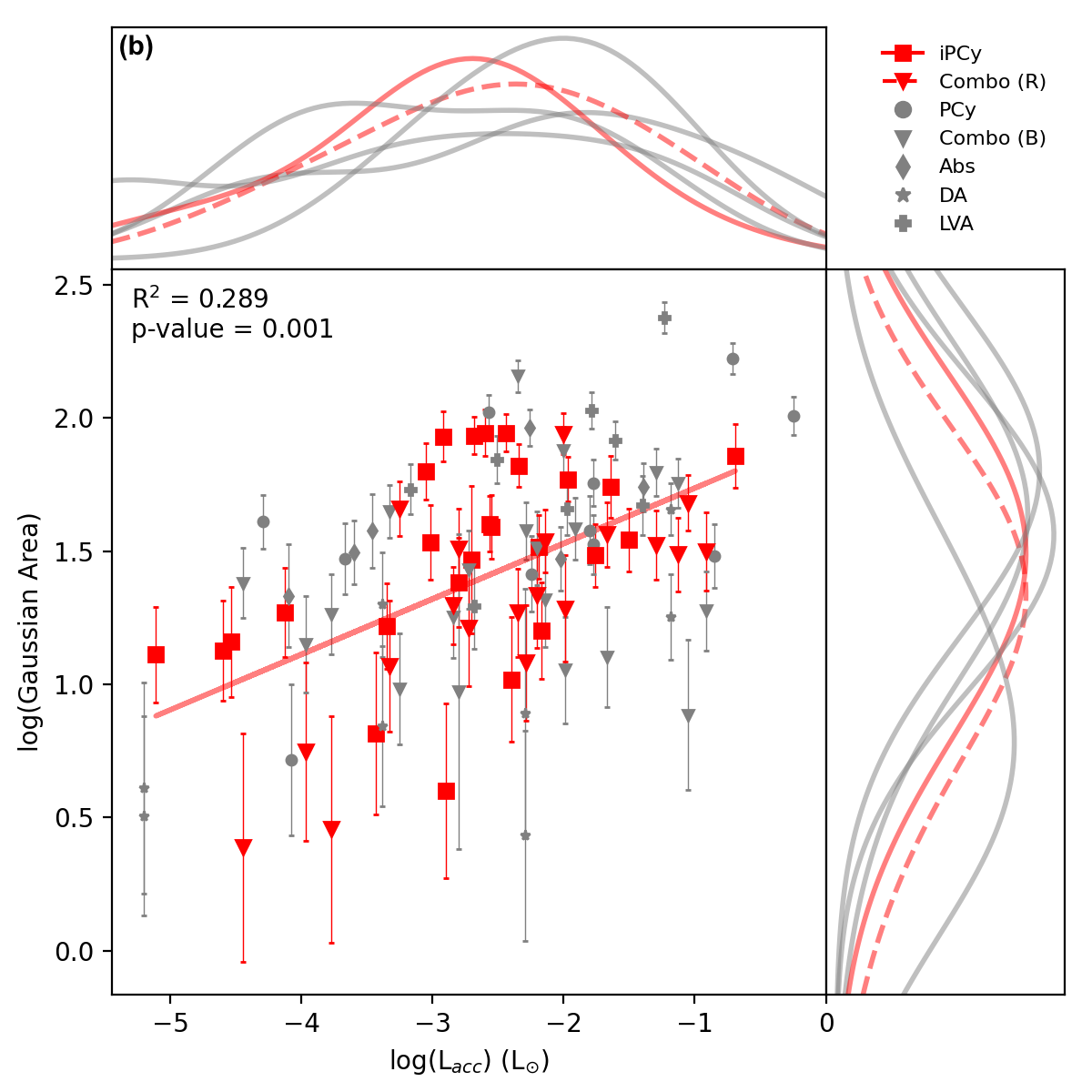}
\caption{Gaussian area, in units of \kms~$\times$~normalised flux units, of the observed absorption features in our spectra, with respect to the accretion luminosity, for the (a) P Cygni and blue combination profile absorption features, and (b) inverse P Cygni and red combination profile absorption features.}
\label{fig::gauss_lacc}
\end{figure}

In Figure \ref{fig::gauss_lacc} we plot the Gaussian area (in units of \kms~$\times$~normalised flux units) of the absorption features in our spectra, which is proportional to the column density of the infalling/outflowing gas, and thus can be related to the mass accretion or mass ejection rates. 
In general, we find that sources with higher accretion luminosities tend to have larger absorption features in the \henir line profile. Given that the red-shifted absorption features trace accretion, we would expect these profiles (i.e. inverse P Cygni and combination) to have higher accretion luminosities than profiles with only blue absorption features. Instead, from the curves along the top of each panel in Figure \ref{fig::gauss_lacc}, we see that P Cygni profiles peak at higher accretion luminosities (log(L$_{\rm acc}$) $\approx$ -2~\lsun) than inverse P Cygni and combination profiles (which peak at (log(L$_{\rm acc}$) $\approx$ -3~\lsun). 

\begin{figure}[h]
    \centering
    \includegraphics[width=0.45\textwidth]{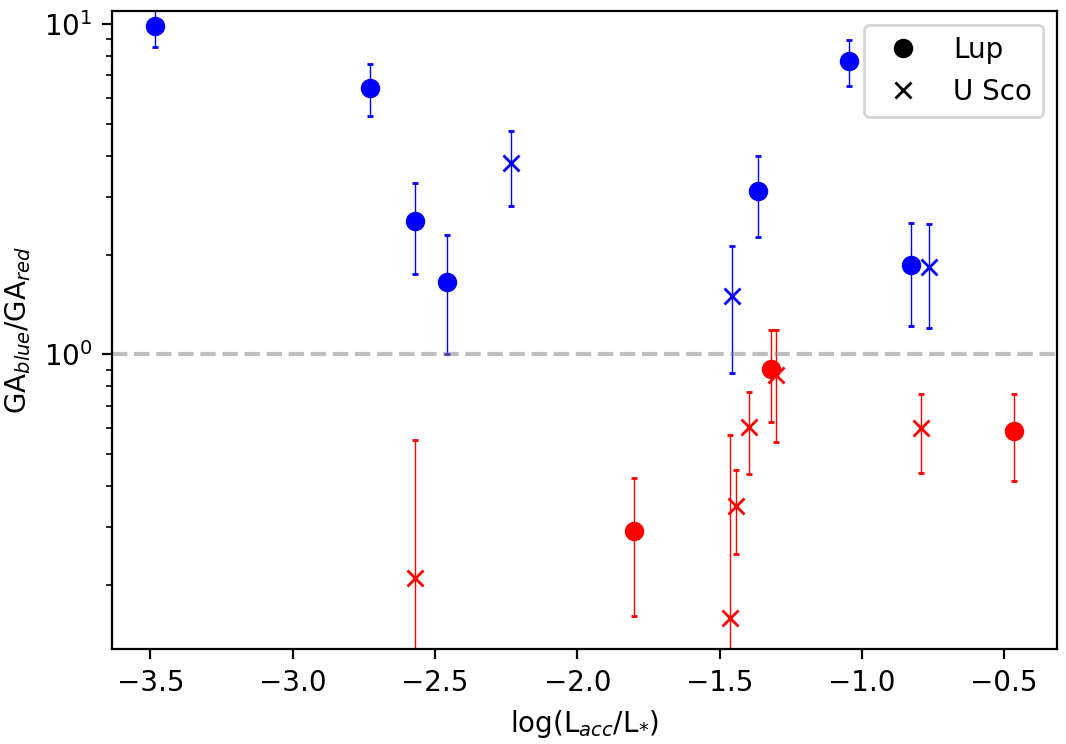}
    \caption{Plot of the Gaussian area, in units of \kms~$\times$~normalised flux units, of the blue-shifted absorption feature divided by that of the red-shifted feature for combination profiles. Combination profiles in Lupus are denoted by circle markers, and those in Upper Sco are marked by an X. The horizontal dashed line (where the ratio is equal to 1) marks where the area of the blue and red features are equal. Blue or red coloured markers indicate blue- or red-dominated sources.}
    \label{fig:combo_area}
\end{figure}

We investigate the combination profiles in more depth in Figure \ref{fig:combo_area}, where we present the ratio of the Gaussian area of the blue-shifted absorption feature divided by that of the red-shifted feature. In this plot there is a trend towards higher accretion luminosities where this ratio is close to or equal to one (dashed line), i.e. when the blue- and red-absorption features are most similar. We note however that the p-value of the linear fits for both the blue and red points was 0.15, i.e. the trend is not statistically significant. This is thought to be due to small number statistics (18 sources in Figure \ref{fig:combo_area}).


\subsection{Evolutionary trends}
In Figure \ref{fig:piecharts}, the proportion of each profile type in both star-forming regions of our sample is presented in comparison to that of the Taurus sample of \citet{edwards06}. Given that the ages of Taurus and Lupus are similar \citep[$<$ 3 Myr][]{luhman10,comeron08}, we would expect the proportions of each profile type in these two regions to be similar. 
However, we note large differences in the proportions of the three most common profile types (P Cygni, inverse P Cygni, and combination profiles) between the Taurus and Lupus regions. As noted previously, this is not thought to be due to instrumental effects. 
It is also important to note that the classification of profile types in the \henir line varies through the literature. For example, \citet{edwards06} define only five distinct profile types, while \citet{thanathibodee2022} identifies six profile types, compared to the eight types we observe. It could be that the double absorption profiles observed in our sample, for example, are in fact combination profiles with no emission feature. In this case, the number of combination profiles in Lupus becomes 16 ($\simeq$ 20\%) compared to 15 ($\simeq$ 39\%) in Taurus. This still does not explain the large difference in the number of P Cygni and inverse P Cygni profiles in each region. However, we note that certain regions of the Lupus and Taurus clouds have varying ages and this may suggest that younger sources with outflow phenomena are more common in Taurus. 

\begin{figure}[h]
    \centering
    \includegraphics[width=0.4\textwidth]{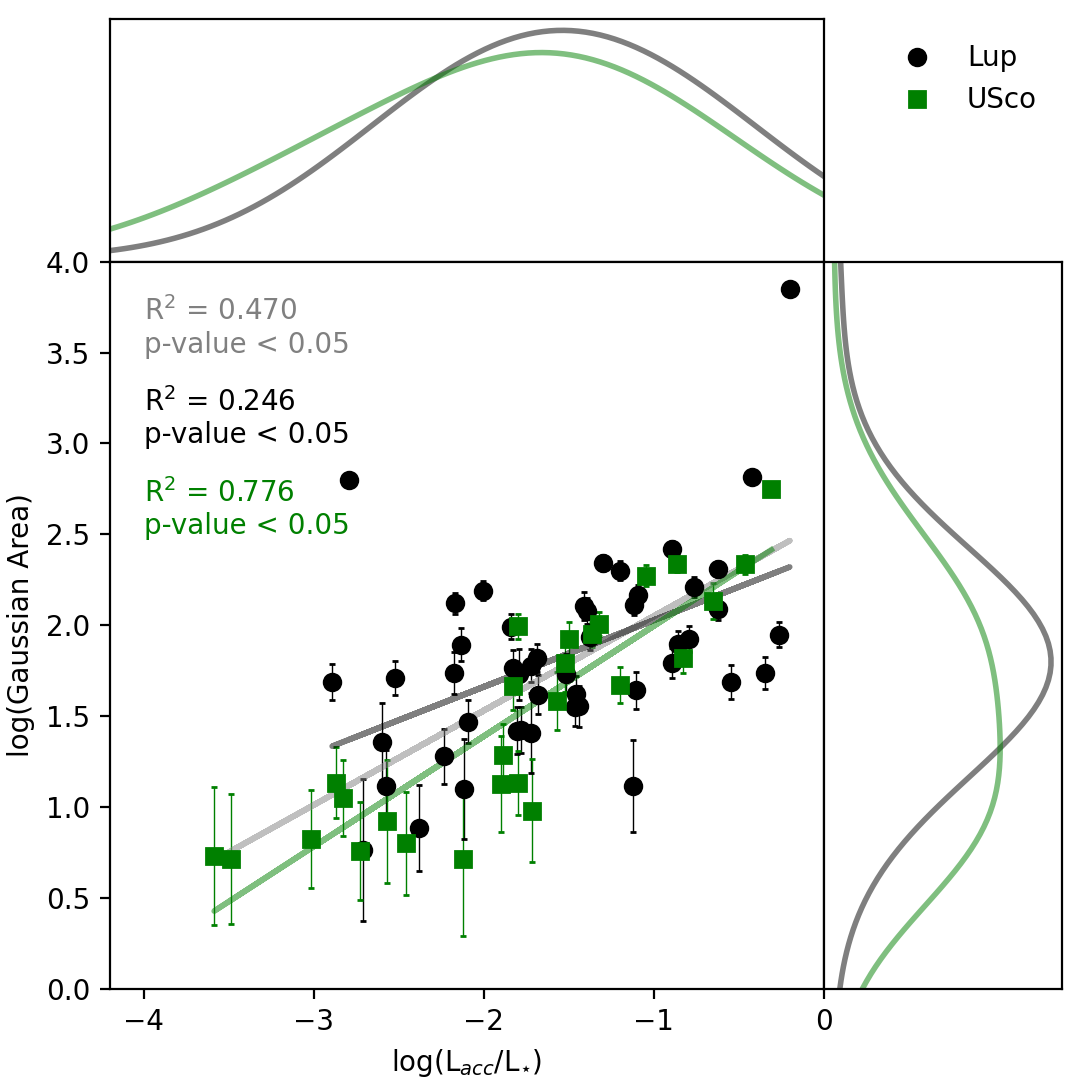}
    \caption{Gaussian area, in units of \kms~$\times$~normalised flux units, versus log(\lacc/L$_{\star}$) of the emission features in pure emission, P Cygni, inverse P Cygni and combination profile types separated by star-forming region. The line of best fit for all datapoints is marked in grey, with a linear fit for Lupus and Upper Sco in black and green, respectively.}
    \label{fig:logGA_em_reg}
\end{figure}

Between the Lupus and Upper Sco regions from our sample, we find two notable differences in the proportions of each of the three most common profile types (P Cygni, inverse P Cygni, and combination profiles). Firstly, the emission-only profiles disappear entirely in our Upper Sco sample. In Section \ref{sect:emonly} we suggest that the emission-only profiles are the result of jets or winds. If this is indeed the case, this suggests that the jet disappears early on, before the age of the Upper Sco sources. In Figure \ref{fig:logGA_em_reg}, we present the Gaussian area, in units of \kms~$\times$~normalised flux units, of the emission features in all emission-only, P Cygni, inverse P Cygni, and combination profiles in both regions. We compare the Gaussian area of the emission features to the ratio of the accretion luminosity to the stellar luminosity for Lupus (black circles) and Upper Sco (green squares). In general, the size of the emission features in Upper Sco is smaller than in Lupus, and only becomes comparable at high L$_{\rm acc}$/L$_{\star}$. In both regions there is a strong trend of the emission with L$_{\rm acc}$/L$_{\star}$ suggesting that at least some of the emission originates from the accretion shocks on the stellar surface. 
Further, this trend supports the results of \citet{kwan2007}, who suggest that stars with higher accretion rates are more likely to exhibit stellar wind signatures.

The second difference between the Lupus and Upper Sco regions in Figure \ref{fig:piecharts} is the significant increase in the proportion of combination profiles, up to $\approx$ 30\% in Upper Sco. This suggests that the sources in Upper Sco are still actively accreting, while the blue-shifted features in the combination profiles means that winds and outflows are also present. In Figure \ref{fig:combo_area}, we also distinguish the ratio of the Gaussian areas of the absorption features in combination profiles between both regions. We find that the majority of the Lupus combination profiles (seven of 10) have a larger blue-shifted feature, while combination profiles in Upper Sco (seven of 9) tend to be dominated by the red-shifted feature, indicating that the winds detected in the Lupus sample are stronger than those in Upper Sco. This also indicates a difference in the accretion in both regions. However, \citet{manara20} found similar values for the mass accretion rates in Lupus and Upper Sco, so this is not likely to be the reason for the increased proportion of combination profiles. We note that the absolute number of combination profiles in Lupus and Upper Sco is similar (12 and 11, respectively) so we cannot rule out the possibility that this difference between the two regions is due to the small sample size.

\section{Conclusions}
\label{sect::concl}

In this work, we present X-Shooter observations of the \henir line from a sample of 107 young stellar objects in the Lupus and Upper Scorpius star forming regions. We characterised the line profiles into one of eight profile types to search for any significant trends in our data with age, source inclination and accretion properties. Our main conclusions are as follows:
\begin{itemize}
\item We observe variations in the proportions of each profile type between the two star forming regions, particularly in the number of profiles with both blue- and red-shifted absorption features. In Lupus, a large proportion of the targets (33\%) have inverse P Cygni profiles, followed by roughly equal proportions of P Cygni (13\%) and combination profiles (15\%). Meanwhile, in Upper Scorpius, the most common profile types are the combination and inverse P Cygni profile types ($\approx$ 33\% each), with fewer P Cygni profiles (28\%) observed. In contrast, \citet{edwards06} observed a larger number of P Cygni profiles and fewer inverse P Cygni profiles in the Taurus sample.

\item The observed maximum velocities of the absorption features are consistent with the expected free-fall, but are larger than the escape velocities of our targets. The free-fall velocities are consistent with the observed velocities in the majority of targets. However, some sources show observed maximum velocities which differ from the expected free-fall velocity which may occur if we are observing material which has not reached the terminal velocity yet, or may be related to the accretion geometry and magnetic field. Generally, the outflowing gas (traced by the blue absorption features), has a velocity larger than the escape velocity, however, magnetic fields associated with MHD disk winds may explain the launching of this material from sources with velocities smaller than the escape velocity. 

\item We find a correlation between the maximum velocities and the source inclination, and this correlation is stronger for sources with blue absorption features than those with only red features. This implies that different velocity components of the wind are observed at different inclinations, as suggested by models of the \henir line \citep{kwan2007}. 

\item The full-width half-maximum of the absorption features show less of a correlation with the source inclination than the velocities, which is important to note if we aim to discriminate between a stellar wind or disk wind. Further, our results are consistent with previous works by \citet{edwards06,kwan2007,fischer2008} where the maximum absorption velocities also appeared to be dependent on the viewing angle. 

\item We typically observe different accretion rates for each profile type in our sample. We see higher accretion rates in targets where the \henir line profile has a blue-shifted absorption feature, supporting the idea that these winds are in fact accretion-powered \citep{edwards06}. This argument is again strengthened when examining the combination profiles, which show the highest accretion rates when the blue and red-shifted absorption features are similar in size. 

\item The observed differences in the proportion of emission-only profiles may suggest that jets or winds are more common in younger sources, and may disappear entirely quite early on (ages $<$ 5 Myr). Meanwhile, the increase in the proportion of combination profiles is not linked to differences in accretion between both regions, and may be due to the smaller sample in Upper Sco.

\end{itemize}

In our study of over 100 sources in Lupus and Upper Sco, we build on previous works by providing statistics for a larger sample of stars than before.  We find that the \henir line is a particularly good tracer of both infalling and outflowing gas in the inner disk. Future studies of even larger samples, and in star-forming regions of different ages, could reveal evolutionary trends in accretion and ejection signatures in young stars. Future work would also benefit from a clearer understanding of inner disk inclinations and any misalignment with the outer disk. Enlarging the statistics on \henir line properties will allow us to confirm our results and past results, and further our understanding of the accretion-ejection connection in the innermost regions of protoplanetary disks. 

\begin{acknowledgements}
We thank the ESO staff in Paranal for carrying out the observations in Service mode. We thank S. Edwards and A. Natta for insightful discussions and feedback on this work, and G. Beccari for the support during this work. This work benefited from discussions with the ODYSSEUS team (HST AR-16129), \url{https://sites.bu.edu/odysseus/}. 
JE acknowledges the ESO Studentship program funded by the Irish Research Council and Teaching Assistantship funding from the School of Physics, University College Dublin. 
Funded by the European Union under the European Union’s Horizon Europe Research \& Innovation Programme 101039452 (WANDA). Views and opinions expressed are however those of the author(s) only and do not necessarily reflect those of the European Union or the European Research Council. Neither the European Union nor the granting authority can be held responsible for them. 
This research received financial support from the project PRIN-INAF 2019 "Spectroscopically Tracing the Disk Dispersal Evolution".
MV acknowledges support from ESA thorugh the Leiden/ESA Astrophysics Program for Summer Students (LEAPS) 2015. We made use of PyAstronomy and Astropy for this work.
\end{acknowledgements}

\bibliography{bibliography}

\onecolumn

\captionsetup{width=0.95\textwidth}
\begin{longtable}{l|cc|cc|ccc}
\caption{Observed profile types, emission centroid velocity and absorption centroid velocities in the Lupus sample. Stellar radii and the calculated escape and free-fall velocities are also listed.}\label{tab:lup_results1}\\
\hline
Object  &   Fit		&	Profile Type	& V$_{em}$  & V$_{abs}$ & R$_{\star}$ & V$_{esc}$ & V$_{ff}$ \\
        &   		&					&	(\kms)				&	(\kms) & (R$_{\odot}$) & (\kms) & (\kms) \\
\hline                                                                      
Sz66                     &      &  iPCy  & -53.88  &  17.53 & 1.34  &    287.09  &    260.29 \\ 
AKC2006-19               &      &  PCy   & 26.97   &  -31.05 & 0.48  &    332.42  &    301.40 \\   
Sz69                     &      &  Em    & -52.28  & & 0.98  &    279.08  &    253.03   \\   
Sz71                     &      &  iPCy  & -49.65  &  98.22 & 1.45  &    328.05  &    297.43 \\   
Sz72                     &   M   &  PCy   & -65.47  &  -196.31 & 1.37  &    321.18  &    291.20 \\   
Sz73                     &  M    &  Em    & -34.23  & & 1.37  &    532.32  &    482.63   \\   
Sz74                     &   M   &  PCy   & 24.32   &  -65.33   & 3.22  &    188.58  &    170.98 \\   
Sz83                     &  M    &  PCy   & -10.53     &  -204.98  & 2.47  &    346.99  &    314.60  \\   
Sz84                     &  M    &  iPCy   &  -48.76    & 16.41  & 1.23  &    236.07  &    214.03  \\   
Sz130                    &      &  Combo  &  2.98    &   -64.98  & 1.12  &    417.30  &    378.35 \\   
						 &		&	   &	  &	170.18 \\
Sz88A                    &  M    &  Em    &  -10.95    &  & 1.58  &    476.99  &    432.47    \\   
Sz88B                    &  M    &  Combo    &  27.41    &  -12.55  & 1.82  &    204.86  &    185.74 \\
						 &		&	   &	  &	175.06 \\   
Sz91                     &  M    &  Combo    &  -30.0    &  -77.2  & 1.09  &    423.01  &    383.53 \\  
						 &		&	   &	  &	164.26 \\ 
Lup713                   &      & iPCy     &  -34.62    &  103.91  & 0.51  &    288.25  &    261.35  \\   
Lup604s                  &      & DA     &      &   -9.26 & 0.71  &    253.17  &    229.54  \\   
						 &		&	   &	  &	113.76 \\
Sz97                     &  M    & iPCy     &  -102.39   &  81.98  & 1.04  &    297.40  &    269.64  \\   
Sz99                     &  M    & iPCy     &  0.0    &  83.52 & 0.70  &    354.57  &    321.48   \\   
Sz100                    &      & Combo    &  -12.52    &  -77.39 & 1.01  &    212.89  &    193.02  \\ 
						 &		&	   &	  &	107.47 \\  
Sz103                    &  M    &  iPCy    &  -114.38    &  -17.67  & 1.08  &    284.87  &    258.28  \\   
Sz104                    &      & Combo     & -2.58     &   -59.24 & 0.90  &    259.82  &    235.57\\   
						 &		&	   &	  &	108.29 \\
Lup706                   &  M    & iPCy     &  -63.64    & 50.0     \\   
Sz106                    &      & Em     & -78.15     &   & 0.57  &    616.00  &    558.50  \\   
Par-Lup3-3               &      & iPCy     &  19.66    & 136.19  & 1.21  &    269.42  &    244.27   \\   
Par-Lup3-4               &      & Em     &  -56.02    &      \\   
Sz110                    &      & Combo     & -13.97     &  -174.80  & 1.32  &    257.41  &    233.39 \\   
						 &		&	   &	  &	132.01 \\
Sz111                    &      & iPCy     &  -109.70    &  76.29  & 1.11  &    417.88  &    378.88 \\ 
Sz112                    &  M    &  Combo    &   -9.89   & -60.22  & 1.18  &    234.05  &    212.21   \\    
						 &		&	   &	  &	91.01 \\
Sz113                    & M     & PCy     &   -10.78   & -147.44     \\   
2MASSJ16085953-3856275   &      & Em     &  -12.30    & &  0.49  &    152.25  &    138.04   \\   
SSTc2d160901.4-392512    &  M    &  iPCy    &   -62.88   &  -10.34  & 0.99  &    298.16  &    270.33  \\   
Sz114                    &  M    &  iPCy    &  22.42    &  80.51  & 1.52  &    218.57  &    198.17  \\   
Sz115                    &      &  Abs    &      &  -15.58  & 1.08  &    265.43  &    240.65  \\   
Lup818s                  &      &  iPCy    &  -24.91    &   85.09 & 0.53  &    240.43  &    217.99 \\   
Sz123A                   & M     &  Em    &  5.59    &  & 0.88  &    543.99  &    493.22    \\    
Sz123B                   & M     &  Em    &  -14.05    &   & 0.46  &    578.41  &    524.42  \\   
SST-Lup3-1               &      &  PCy    &  72.55    &   108.47 & 1.71  &    189.00  &    171.36  \\   
Sz65                     &   & PCy   & 19.81  &  -439.49 & 1.91  &    373.91  &    339.01 \\
AKC2006-18               &   & iPCy   & 3.02  & 133.01 & 0.39  &    262.54  &    238.03  \\
SSTc2dJ154508.9-341734   &   & PCy   & -40.24  & -127.07 & 0.87  &    238.34  &    216.09 \\
Sz68                     &   & DA   &   &   -141.33  & & &   \\
						 &   &    &   &   15.77  & & &   \\
SSTc2dJ154518.5-342125   &   &  PCy  & -15.97  &  -66.31  & 0.77  &    252.99  &    229.37 \\
Sz81A                    &   &  Abs  &   &  -100.49 & 1.03  &    291.78  &    264.54 \\
Sz81B                    &   &  DA  &   &  -125.68 & 1.67  &    185.01  &    167.74 \\
						 &   &    &   &   15.29  & & &   \\
Sz129                    &   & Combo   &  0.60 & -103.65 & 1.33  &    533.35  &    483.57  \\
						 &   &    &   &   212.72 & & &   \\
SSTc2dJ155925.2-423507   &   & Combo   &  6.07 & -134.31  & 0.48  &    332.42  &    301.40 \\
						 &   &    &   &   143.03 & & &   \\
RY\,Lup                  &   &  PCy  & 36.31  & -72.47  & & &   \\ 
SSTc2dJ160000.6-422158   &   &  iPCy  & -128.615  & -38.28 & 0.65  &    358.82  &    325.33 \\
SSTc2dJ160002.4-422216   &  M &  Abs  &   &    86.80  & 1.32  &    257.41  &    233.39 \\
SSTc2dJ160026.1-415356   &   &  iPCy  & 3.75  &  117.0  & 1.01  &    230.17  &    208.69  \\
MY\,Lup                  &   &  Abs  &   &   -43.01  & 1.18  &    565.02  &    512.29 \\ 
Sz131                    & M  &  iPCy  & -71.51  &  13.46  & 1.11  &    321.34  &    291.35  \\
Sz133                    &   &  Em  &  -88.30 &  & & &    \\
SSTc2dJ160703.9-391112   &   &  Abs  &   & -11.43 & & &  \\
Sz90                     & M  &  LVA  & -8.88  &  8.86 & 1.31  &    568.08  &    515.06 \\
Sz95                     & M  &  Combo  & -108.51  & -13.03   & 1.46  &    346.79  &    314.42  \\
						 &   &    &   &   100.0 & & &   \\
Sz96                     &  M &  LVA  & -24.56  & 2.72 & 1.58  &    330.08  &    299.27 \\
2MASSJ16081497-3857145   & M  &  Abs  &   & 41.41 & 0.36  &    327.16  &    296.62 \\
Sz98                     &  M &  PCy  & -46.63  & -92.88  & 2.50  &    456.82  &    414.18  \\
Lup607                   &  M &  Em  & 1.52  &  & 0.87  &    199.08  &    180.50 \\
Sz102                    &   &  Em  &  -8.23 &   & & &   \\ 
SSTc2dJ160830.7-382827   &   &  Abs  &   &  -16.80  & & &   \\ 
SSTc2dJ160836.2-392302  & M  &  Misc  &  1.49 &  -180.90 & 2.02  &    511.84  &    464.07 \\
						 &   &    &   &   -12.24 & & &   \\
						 &   &    &   &   100.0 & & &   \\
Sz108B                   & M  &  Combo  & -80.37  & 10.97 & 1.13  &    239.20  &    216.87 \\
						 &   &    &   &   100.0 & & &   \\
2MASSJ16085324-3914401   &   &  Combo  & 41.55  &  -33.77  & 1.31  &    290.45  &    263.34 \\
						 &   &    &   &   150.35 & & &   \\
2MASSJ16085529-3848481   &   &  iPCy  &  16.60 & 115.96  & 0.87  &    199.08  &    180.50 \\
SSTc2dJ160927.0-383628   &   &  iPCy  & -71.56  &  23.62 & 0.86  &    318.99  &    289.21 \\
Sz117                    &   &  iPCy  & -47.30  & 123.37 & 1.58  &    245.45  &    222.54 \\
Sz118                    &   &  iPCy  &  -89.95 & 28.03 & 1.50  &    514.88  &    466.82 \\
2MASSJ16100133-3906449   &   &  iPCy  &  45.07 & 133.03 & & &   \\
SSTc2dJ161018.6-383613   &   &  Abs  &   & 22.73  & 0.68  &    289.34  &    262.34 \\
SSTc2dJ161019.8-383607   & M  &  Abs  &   & 42.15  & 0.77  &    221.88  &    201.17 \\
SSTc2dJ161029.6-392215   &   &  DA  &   &  -24.37  & 1.08  &    265.68  &    240.88 \\
						 &   &    &   &   126.37 & & &   \\
SSTc2dJ161243.8-381503   &   &  iPCy  & -70.65  &  21.68  & 1.52  &    336.25  &    304.87  \\
SSTc2dJ161344.1-373646   & M  &  PCy  & 1.52  & -78.32  & 0.68  &    308.03  &    279.28 \\

Sz75/GQ\,Lup             & M  &  Combo  & -85.14  &  0.0 & 2.30  &    364.57  &    330.54  \\
						 &   &    &   &   161.403 & & &   \\
Sz76                     &   & Misc  &  33.56 &  -108.97 & 1.32  &    278.90  &    252.87 \\
						 &   &    &   &   -20.24 & & &   \\
						 &   &    &   &   132.09 & & &   \\
Sz77                     &   &  Combo  & -18.51  &  -117.90  & 1.56  &    428.92  &    388.89 \\
						 &   &    &   &   125.19 & & &   \\
RXJ1556.1-3655           &   &  Em  &  22.10 &   & 1.24  &    388.31  &    352.07  \\
Sz82/IM\,Lup             &   &  Combo  & -7.86  &  -97.27 & 2.84  &    356.98  &    323.66 \\
						 &   &    &   &   132.78 & & &   \\ 
EX\,Lup                  &   &  Combo  & 0.736  &  -134.77 & 1.96  &    320.94  &    290.993 \\
						 &   &    &   &   217.083 & & &   \\ 
\hline
\end{longtable}
\tablefoot{An 'M' in column 2 is spectra that was fit using manually defined initial guesses}

\begin{longtable}{l|cc|cc|ccc}
\caption{Observed profile types, emission centroid velocity and absorption centroid velocities in the Upper Scorpius sample. Stellar radii and the calculated escape and free-fall velocities are also listed. }\label{tab:usco_results1}\\
\hline
Object  &   Fit		&	Profile Type	& V$_{em}$  & V$_{abs}$ & R$_{\star}$ & V$_{esc}$ & V$_{ff}$ \\
        &   		&					&	(\kms)				&	(\kms) & (R$_{\odot}$) & (\kms) & (\kms) \\
\hline                                                                      
2MASSJ15534211-2049282     &     &  iPCy  & -87.78   & 23.79 & 0.99   &   304.57  &    276.14 \\
2MASSJ15583692-2257153     &  M   &  PCy  & 11.34   & -80.22 &   1.79   &   589.50  &    534.48 \\
2MASSJ16001844-2230114     &  M   &  PCy  & 45.14   & -274.83 &  0.98   &   279.34  &    253.27 \\
2MASSJ16035767-2031055     &     &  PCy  & 11.78   & -69.55 &   1.33   &   510.01  &    462.41 \\
2MASSJ16035793-1942108     &     &  DA  &    & 14.11 &  0.87   &   428.31  &    388.34 \\
						   &     &    &    & 144.62  \\
2MASSJ16041740-1942287     &     &  Combo  &   71.3 & -12.68 & 0.95   &   352.99  &    320.04 \\
						   &     &    &    & 150.68  \\
2MASSJ16041893-2430392     &     &  Combo  &  48.95 & -9.45 & 1.77   &   282.67  &    256.29  \\
						   &     &    &    & 157.44  \\
2MASSJ15354856-2958551\_E  &     &  iPCy  & -11.78   & 106.82 &    1.03   &   272.08  &    246.69 \\
2MASSJ15354856-2958551\_W  &     &  iPCy  & -0.33   & 126.56 &  1.03   &   272.08  &    246.69 \\
2MASSJ15514032-2146103     & M    & Abs   &    &  49.37 &    0.73   &   315.37  &    285.93 \\
2MASSJ15530132-2114135     &     &  iPCy  & -55.62   & 118.21 &   0.73   &   315.37  &    285.93 \\
2MASSJ15582981-2310077     & M    &  Combo  &   -27.6 & -100.0 & 0.73   &   315.37  &    285.93 \\
						   &     &    &    & 101.87  \\
2MASSJ16014086-2258103     &  M   &  LVA  &  -10.73  &  -27.51 &  0.99   &   345.39  &    313.15 \\
2MASSJ16020757-2257467     &     &  Combo  &   35.15 & -44.86 &  0.79   &   460.95  &    417.92 \\
						   &     &    &    & 137.56  \\
2MASSJ16024152-2138245     &     &  Combo  &   -6.25 & -124.45 &  0.62   &   272.31  &    246.89 \\
						   &     &    &    & 124.24  \\
2MASSJ16054540-2023088     &     &  Combo  &   1.66 & -34.35 &   1.03   &   272.08  &    246.69 \\
						   &     &    &    & 124.78  \\
2MASSJ16062196-1928445     &     &  Combo  &   34.11 & -73.48 &  1.42   &   351.83  &    318.99  \\
						   &     &    &    & 191.51  \\
2MASSJ16063539-2516510     & M    &  iPCy  &  -78.39  & -5.61 &    0.56   &   348.77  &    316.22 \\
2MASSJ16064385-1908056     &     &  DA  &    & -78.86 & 1.09   &   535.63  &    485.64 \\
						   &     &    &    & 118.42  \\
2MASSJ16072625-2432079     &     & iPCy   & -56.43   & 70.64 &    1.25   &   297.81  &    270.01 \\
2MASSJ16081566-2222199     &     & PCy   &  48.59  & -8.53 &   1.00   &   395.08  &    358.21 \\
2MASSJ16082324-1930009     & M    &  LVA  &  -27.51  & 13.66 &  1.33   &   414.53  &    375.84 \\
2MASSJ16082751-1949047     & M    & iPCy   &  -64.12  &  0.0 &  0.87   &   247.33  &    224.25 \\
2MASSJ16090002-1908368     &     & iPCy   &  -43.78  &  115.16 &  0.73   &   315.37  &    285.93 \\
2MASSJ16090075-1908526     & M    & iPCy   &  -57.09  &  50.0 &   1.35   &   411.62  &    373.20 \\
2MASSJ16095361-1754474     &  M   & iPCy   &  -72.74  &  126.61 &   0.56   &   348.77  &    316.22 \\
2MASSJ16104636-1840598     &     & Combo  &   27.89 & -67.28 &  0.65   &   333.46  &    302.34 \\
						   &     &    &    & 114.3  \\
2MASSJ16111330-2019029     &     & Combo  &   -23.26 & -177.56 &    0.52   &   445.84  &    404.23 \\
						   &     &    &    & 123.46  \\
2MASSJ16123916-1859284     &  M   &  PCy  & 0.0   & -62.801 &    1.19   &   400.18  &    362.83  \\
2MASSJ16133650-2503473     &  M   &  LVA  & -13.8   & -30.13 &   0.95   &   358.63  &    325.16 \\
2MASSJ16135434-2320342     &  M   &  Combo  &   -12.73 & -25.74 &  1.13   &   259.96  &   235.69  \\
						   &     &    &    & 136.06  \\
2MASSJ16141107-2305362     &     &  Abs  &    & -48.91  &   1.62   &   542.01  &    491.42 \\
2MASSJ16143367-1900133     &     &  Combo  &   -38.24 & -137.23 &  2.10   &   229.36  &    207.96 \\
						   &     &    &    & 68.81  \\
2MASSJ16154416-1921171     & M    & iPCy   &  -115.87  & 59.07 &  1.21   &   504.34  &    457.27 \\
2MASSJ16181904-2028479     & M    &  iPCy  &  -80.19  & -18.31 &  0.84   &   270.03  &    244.82  \\
\hline
\end{longtable}
\tablefoot{An 'M' in column 2 is spectra that was fit using manually defined initial guesses}

\begin{appendix}
\section{Tables of source information}

\captionsetup{width=0.95\textwidth}
\begin{longtable}{l|c|cc|c|cccc|l}
\caption{Stellar and accretion properties for the targets in the Lupus region from \citet{frasca2017,alcala19}.}\label{tab:lup_properties1}\\
\hline
Object  &   dist & SpT & $A_{\rm V}$   & i$_{disk}$ & \lstar & log\lacc & \mstar & log\macc & Notes \\
        &   [pc] &     & [mag.]        & [$^{\circ}$] & [\lsun]      &   & [\msun ]      &  &  \\
\hline                                                                      
Sz66                     &  157 & M3  &  1.00 &  69.0   & 0.22  &  -1.76  &  0.29 &   -8.50 & LVC        \\
AKC2006-19               &  153 & M5  &  0.00 &  60.0   & 0.02  &  -4.08  &  0.14 &  -10.97 &         \\
Sz69                     &  155 & M4.5 & 0.00 &  43.53   & 0.09  &  -2.77  &  0.20 &   -9.48 & LVC, HVC        \\
Sz71                     &  156 & M1.5 & 0.50 &  40.82   & 0.33  &  -2.17  &  0.41 &   -9.02 &  LVC       \\
Sz72                     &  156 & M2  &  0.75 &  75.0   & 0.27  &  -1.77  &  0.37 &   -8.60 & LVC, HVC        \\
Sz73                     &  157 & K7  &  3.50 &  49.76   & 0.46  &  -0.96  &  0.78 &   -8.12 & LVC, HVC        \\
Sz74                     &  159 & M3.5 & 1.50 &     & 1.16  &  -1.45  &  0.30 &   -7.81 &  LVC       \\
Sz83                     &  160 & K7  &  0.00 &  18.0   & 1.49  &  -0.25  &  0.67 &   -7.08 &         \\
Sz84                     &  153 & M5  &  0.00 & 73.99    & 0.13  &  -2.68  &  0.17 &   -9.23 &  td, LVC     \\ 
Sz130                    &  160 & M2  &  0.00 &  55.0   & 0.18  &  -2.14  &  0.39 &   -9.09 &  LVC, HVC       \\
Sz88A                    &  158 & M0  &  0.25 &  50.0   & 0.31  &  -1.40  &  0.61 &   -8.49 &  LVC       \\
Sz88B                    &  159 & M4.5 & 0.00 &  50.0   & 0.07  &  -3.30  &  0.20 &  -10.05 &  LVC      \\
Sz91                     &  159 & M1 &  1.20  &  48.0   & 0.20  &  -2.00  &  0.51 &   -9.07 &  td, LVC   \\ 
Lup713                   &  174 & M5.5 & 0.00 &     & 0.02  &  -3.62  &  0.11 &  -10.40 &         \\
Lup604s                  &  160 & M5.5 & 0.00 &  0.0   & 0.04  &  -3.89  &  0.12 &  -10.56 &  LVC       \\
Sz97                     &  158 & M4  &  0.00 &  0.0   & 0.11  &  -3.11  &  0.24 &   -9.88 &  LVC       \\
Sz99                     &  159 & M4  &  0.00 &     & 0.05  &  -2.80  &  0.23 &   -9.73 &  LVC, HVC       \\
Sz100                    &  137 & M5.5 & 0.00 &  45.11   & 0.08  &  -3.33  &  0.14 &   -9.87 &  td, LVC, HVC     \\ 
Sz103                    &  160 & M4 &  0.70  &  50.0   & 0.12  &  -2.60  &  0.23 &   -9.33 &  LVC, HVC       \\
Sz104                    &  166 & M5  &  0.00 &  0.0   & 0.07  &  -3.36  &  0.16 &  -10.03 &  LVC       \\
Lup706                   &  159 & M7.5 & 0.00 &    & 0.002 &  -5.00  &  0.06 &  -11.90 &  sl     \\
Sz106                    &  162 & M0.5 & 1.00 &     & 0.06  &  -2.68  &  0.57 &  -10.07 &  sl, LVC       \\
Par-Lup3-3               &  159 & M4  &  2.20 &  0.0   & 0.15  &  -3.10  &  0.23 &   -9.77 &         \\
Par-Lup3-4               &  151 & M4.5 & 0.00 &     & 0.002 &  -4.35  &  0.17 &  -11.81 &  sl, LVC   \\
Sz110                    &  160 & M4   & 0.00 &  63.0   & 0.18  &  -2.20  &  0.23 &   -8.84 &  LVC    \\
Sz111                    &  158 & M1  &  0.00 &  55.0   & 0.21  &  -2.40  &  0.51 &   -9.47 &  td, LVC   \\ 
Sz112                    &  160 & M5  &  0.00 &  0.0   & 0.12  &  -3.39  &  0.17 &   -9.94 &  td, LVC     \\ 
Sz113                    &  163 & M4.5 & 1.00 &  10.78   & 0.04  &  -2.28  &  0.19 &   -9.12 & LVC, HVC    \\
2MASSJ16085953-3856275   &  150 & M8.5 & 0.00 &     & 0.01  &  -4.85  &  0.02 &  -11.02 & LVC     \\
SSTc2d160901.4-392512    &  164 & M4  &  0.50 &  60.0   & 0.10  &  -3.17  &  0.23 &   -9.95 & LVC    \\
Sz114                    &  162 & M4.8 & 0.30 &  15.84   & 0.21  &  -2.68  &  0.19 &   -9.17 & LVC, HVC      \\
Sz115                    &  158 & M4.5 & 0.50 &     & 0.11  &  -2.91  &  0.20 &   -9.57 &     \\
Lup818s                  &  157 & M6  &  0.00 &  0.0   & 0.02  &  -4.31  &  0.08 &  -10.96 & LVC     \\
Sz123A                   &  159 & M1  &  1.25 &  40.0   & 0.13  &  -2.00  &  0.55 &   -9.21 & td, LVC, HVC   \\ 
Sz123B                   &  159 & M2  &  0.00 &  40.0   & 0.03  &  -2.90  &  0.40 &  -10.24 & sl, LVC, HVC       \\
SST-Lup3-1               &  165 & M5  &  0.00 &  73.0   & 0.04  &  -3.77  &  0.16 &  -10.53  &        \\
Sz65                     &  155 & K7  &  0.60 &  61.46   & 0.89  &  -2.57      &  0.70 &   -9.52 &      \\
AKC2006-18               &  139 & M6.5 & 0.00 &  60.0   & 0.01  &  -4.60      &  0.07 &  -11.23 &  LVC     \\
SSTc2dJ154508.9-341734   &  155 & M5.5 & 5.50 &  47.0   &  0.06  &  -1.77     &  0.14 &   -8.36 &  LVC, HVC  \\
Sz68                     &  154 & K2  &  1.00 &  32.89   & 5.42 & -1.18 & 2.15    & -8.39   &     \\
SSTc2dJ154518.5-342125   &  152 & M6.5 & 0.00 &  25.8   & 0.04  &  -4.29     &  0.08 &  -10.72 &  LVC, HVC     \\
Sz81A                    &  160 & M4.5 & 0.00 &  0.0   & 0.25  &  -2.44     &  0.19 &   -8.92 &  LVC    \\
Sz81B                    &  160 & M5.5 & 0.00 &  0.0   & 0.12  &  -3.14     &  0.15 &   -9.61 &  LVC     \\
Sz129                    &  162 & K7  &  0.90 &  31.74   & 0.43  &  -1.13      &  0.78 &   -8.30  &   LVC   \\
SSTc2dJ155925.2-423507   &  147 & M5  &  0.00 &     & 0.02  &  -4.42      &  0.14 &  -11.29  & LVC      \\
RY\,Lup                  &  159 & K2  &  0.40 &  55.0   & 1.87 & -0.85 & 1.53    & -8.16   & td, LVC   \\ 
SSTc2dJ160000.6-422158   &  161 & M4.5 & 0.00 &  0.0   & 0.10  &  -3.04     &  0.20 &   -9.73 & LVC  \\
SSTc2dJ160002.4-422216   &  164 & M4  & 1.40  &  66.0   & 0.18  &  -2.92      &  0.23 &   -9.56  &      \\
SSTc2dJ160026.1-415356   &  164 & M5.5 & 0.90 &     & 0.08  &  -3.22      &  0.14 &   -9.76 &       \\
MY\,Lup                  &  157 & K0  &  1.30 &  72.98  & 0.85  & -0.65 &  1.09 & -8.01 & td       \\ 
Sz131                    &  160 & M3  &  1.30 &  59.0   & 0.15  &  -2.34      &  0.30 &   -9.18  &  LVC, HVC     \\
Sz133                    &  153 & K5 &  1.80  &  78.53   & 0.07  &  -1.78   &        &    &  sl, bz, LVC   \\
SSTc2dJ160703.9-391112   &  159 & M4.5 & 0.60 &  61.0   & 0.003 &  -5.40   & 0.16   &  -12.76   & sl, td?, LVC  \\
Sz90                     &  160 & K7  &  1.80 &   61.31  & 0.42  &  -1.79      &  0.78 &   -8.96  & LVC, HVC   \\
Sz95                     &  158 & M3  &  0.80 &  50.0   & 0.26  &  -2.70      &  0.29 &   -9.40 &    \\
Sz96                     &  157 & M1  &  0.80 &  50.0   & 0.42  &  -2.51      &  0.45 &   -9.37 &  LVC  \\
2MASSJ16081497-3857145  & 159 & M5.5  &  1.50 &  75.0   & 0.01  &  -3.60      &  0.10 &  -10.60 &   LVC  \\
Sz98                     &  156 & K7  &  1.00 &  47.10   & 1.53  &  -0.71      &  0.67 &   -7.54 &  LVC, HVC  \\
Lup607                   &  175 & M6.5 & 0.00 &  0.0   & 0.05  &  -5.02      &  0.09 &  -11.60 &  LVC  \\
Sz102                    &  159 & K2  &  0.70 &  57.60   & 0.01 & -2.20  &    &   & sl, bz, LVC, HVC   \\ 
SSTc2dJ160830.7-382827   &  156 & K2  &  0.20 &  73.0   & 1.84 & -2.02 & 1.53    &  -9.32   &  td, LVC       \\ 
SSTc2dJ160836.2-392302   &  154 & K6m & 1.70  &  56.2   &  1.15  &  -1.03     &  0.83 &   -8.04 & td?, LVC, HVC  \\
Sz108B                   &  169 & M5  &  1.60 &  49.09   & 0.11  &  -3.05      &  0.17 &   -9.62  &  LVC, HVC \\
2MASSJ16085324-3914401   &  168 & M3  &  1.90 &  60.72   & 0.21  &  -3.25      &  0.29 &  -10.00 &     \\
2MASSJ16085373-3914367   &  159 & M5.5 & 4.00 &     & 0.00  &  -3.90      &  0.10 &  -10.94 &     \\
2MASSJ16085529-3848481   &  158 & M6.5 & 0.00 &  0.0   & 0.05  &  -4.31      &  0.09 &  -10.72 &  LVC, HVC  \\
SSTc2dJ160927.0-383628   &  159 & M4.5 & 2.20 &  51.0   & 0.07  &  -1.50      &  0.20 &   -8.25 &  LVC, HVC   \\
Sz117                    &  159 & M3.5 & 0.50 &  0.0   & 0.28  &  -2.30      &  0.25 &   -8.91 & LVC, HVC  \\
Sz118                    &  164 & K5  &  1.90 &  67.0   & 0.72  &  -1.97      &  1.04 &   -9.21 &  LVC   \\
2MASSJ16100133-3906449   &  193 & M6.5 & 1.70 &  35.0   & 0.19  &  -3.43 &  0.14   & -9.74   &  LVC \\
SSTc2dJ161018.6-383613   &  159 & M5  &  0.50 &     & 0.04  &  -4.00      &  0.15 &  -10.76 &   \\
SSTc2dJ161019.8-383607   &  159 & M6.5 & 0.00 &  55.0   & 0.04  &  -4.10      &  0.08 &  -10.52 &    \\
SSTc2dJ161029.6-392215   &  163 & M4.5 & 0.90 &  66.54   & 0.11  &  -3.38      &  0.20 &  -10.05 & td, LVC      \\
SSTc2dJ161243.8-381503   &  160 & M1  &  0.80 &  43.69   & 0.39  &  -2.19      &  0.45 &   -9.07 &   LVC     \\
SSTc2dJ161344.1-373646   &  160 & M5  & 0.60  &  0.0   & 0.04  &  -2.49      &  0.16 &   -9.24 &   LVC, HVC     \\
Sz75/GQ\,Lup             & 152 & K6  &  0.70  &  60.60   & 1.48  &  -0.69      &  0.80 &   -7.63 &  LVC   \\
Sz76                     & 160 & M4  &  0.20  &  38.9   & 0.18  &  -2.55      &  0.23 &   -9.18 & td, LVC        \\
Sz77                     & 155 & K7   &  0.00 &  61.70   &  0.59  &  -1.67      &  0.75 &   -8.76  &    \\
RXJ1556.1-3655           & 158 & M1   &  1.00 &  49.4   & 0.26  &  -0.85      &  0.49 &   -7.85 &  LVC     \\
Sz82/IM\,Lup             & 159 & K5   &  0.90 &  48.40   & 2.60  &  -1.05      &  0.95 &   -7.98 &  td       \\ 
EX\,Lup                  & 158 & M0   &  1.10 &  30.80   & 0.76  &  -0.91      &  0.53 &   -7.74 &    \\
 \hline
\end{longtable}
\tablefoot{Disk inclinations from \citet{tazzari17,ansdell18,yen2018}. Information on the presence of winds or jets is from optical emission lines analysed in \citet{natta14} and \citet{nisini18} \\ td : YSO with transitional disk ; sl : sub-luminous YSO; bz : sub-luminous object falling below the zero-age main-sequence on the HR diagram }

\begin{longtable}{l|c|cc|c|cccc|l} 
\caption{Stellar and accretion properties for the targets in the Upper Scorpius region from \citet{manara20}.}\label{tab:usco_properties}\\
\hline
Object  &  dist & SpT & $A_{\rm V}$   & i$_{disk}$ & \lstar & log\lacc & \mstar & log\macc & Notes \\
        &   [pc] &     & [mag.]        & [$^{\circ}$] & [\lsun]      &   & [\msun ]      &  &  \\
\hline
2MASSJ15534211-2049282   & 136  $\pm$  4 &    M4   & 1.2 &  89.0   &  0.09 & -2.6 &  0.24  &   3.66$\cdot10^{-10}$   &  f  \\
2MASSJ15583692-2257153  & 166  $\pm$  4 &    K0   & 0.0 &     &  2.57 & -0.5 &  1.63$^*$  &   1.59$\cdot10^{-08}$  & f \\
2MASSJ16001844-2230114    & 138  $\pm$  9 & M4.5   & 0.8 &  45.0   &  0.08 & -1.9 &  0.20  &   2.03$\cdot10^{-09}$   & f \\
2MASSJ16035767-2031055  & 143  $\pm$  1 &    K6   & 0.7 &  69.0   &  0.48 & -1.8 &  0.91  &   8.81$\cdot10^{-10}$   &  f \\
2MASSJ16035793-1942108   & 158  $\pm$  2 &    M2    &  0.3 &  56.0   &  0.13 & -5.1 &  0.42  &   6.69$\cdot10^{-13}$  &  f  \\
2MASSJ16041740-1942287    & 161  $\pm$  2 &    M3    & 0.7 & 80.0    &  0.14 & -4.3 &  0.31  &   6.04$\cdot10^{-12}$  &  f \\
2MASSJ16041893-2430392   & 145   & M2   &  0.3 &     &  0.45 & -3.1 &  0.37  &   1.48$\cdot10^{-10}$  &    \\
2MASSJ16042165-2130284       & 150  $\pm$  1 &   K3   & 1.4 &     &  0.90 & -3.2 &  1.24  &   3.09$\cdot10^{-11}$  &  td  \\
2MASSJ15354856-2958551\_E    & 145   & M4.5 & 0.0 &  46.0   &  0.10 & -2.8 &  0.20  &   3.53$\cdot10^{-10}$  & f,b \\
2MASSJ15354856-2958551\_W   & 145   & M4.5 &  0.0 &  46.0   &  0.10 & -2.9 &  0.20  &   2.73$\cdot10^{-10}$  &  b  \\
2MASSJ15514032-2146103       & 142  $\pm$  2 &   M4.5  &  0.3 &  83.0   &  0.05 & -3.5 &  0.19  &   5.01$\cdot10^{-11}$  &  e  \\
2MASSJ15530132-2114135      & 146  $\pm$  2 &   M4.5  & 0.8 &   47.0  &  0.05 & -3.0 &  0.19  &   1.52$\cdot10^{-10}$ &  f  \\
2MASSJ15582981-2310077    & 147  $\pm$  3 &   M4.5  & 1.0 &  32.0   &  0.05 & -2.3 &  0.19  &   7.16$\cdot10^{-10}$  & f \\
2MASSJ16014086-2258103   & 145   & M3   &  1.2 &  74.0   &  0.12 & -1.2 &  0.31  &   7.42$\cdot10^{-09}$   &   f \\
2MASSJ16020757-2257467   & 140  $\pm$  1 &   M2    & 0.4 &  57.0   &  0.08 & -3.8 &  0.44  &   1.08$\cdot10^{-11}$  & f \\
2MASSJ16024152-2138245  & 142  $\pm$  2 &   M5.5  & 0.6 &  41.0   &  0.03 & -2.9 &  0.12  &   2.76$\cdot10^ {-10}$   & f   \\
2MASSJ16054540-2023088  & 145  $\pm$  2 &   M4.5  & 0.6 &  67.0   &  0.10 & -2.8 &  0.20  &   3.58$\cdot10^{-10}$   & f  \\
2MASSJ16062196-1928445   & 145   & M1 & 0.8 &  85.0   &  0.34 & -1.3 &  0.46  &   6.13$\cdot10^{-09}$  & td  \\
2MASSJ16063539-2516510    & 139  $\pm$  3 &   M4.5   & 0.0 &  74.0   &  0.03 & -5.1 &  0.18  &   8.62$\cdot10^{-13}$  &  e \\
2MASSJ16064385-1908056   & 144  $\pm$  7 &   K7   & 0.4 &  48.0   &  0.29 & -2.3 &  0.82  &   2.65$\cdot10^{-10}$ &  e  \\
2MASSJ16072625-2432079   & 143  $\pm$  2 &   M3   & 0.7 &  43.0   &  0.18 & -2.6 &  0.29  &   4.56$\cdot10^{-10}$  &  f \\
2MASSJ16081566-2222199   & 140  $\pm$  2 &   M2   & 0.5 &   86.0  &  0.15 & -3.7 &  0.41  &   1.99$\cdot10^{-11}$  &  f \\
2MASSJ16082324-1930009    & 138  $\pm$  1 &   M0   & 1.1 &  74.0   &  0.32 & -2.0 &  0.61  &   7.90$\cdot10^{-10}$  & f  \\
2MASSJ16082751-1949047   & 145   & M5.5  & 0.6 &  41.0   &  0.06 & -3.1 &  0.14  &   1.97$\cdot10^{-10}$  &  e \\
2MASSJ16090002-1908368   & 139  $\pm$  3 &   M4.5  & 0.3 &  63.0   &  0.05 & -4.2 &  0.19  &   1.02$\cdot10^{-11}$  &  f \\
2MASSJ16090075-1908526  & 138  $\pm$  1 &   M0    & 1.0 &   56.0   & 0.32 & -1.7 &  0.60  &   1.74$\cdot10^{-09}$  & f  \\
2MASSJ16095361-1754474    & 158  $\pm$  5 &   M4.5  & 0.5 &  86.0   &  0.04 & -4.5 &  0.18  &   4.54$\cdot10^{-12}$  & f  \\
2MASSJ16104636-1840598     & 143  $\pm$  3 &   M4.5   & 1.2 &   71.0   & 0.04 & -3.9 &  0.19  &   1.45$\cdot10^{-11}$  &  f \\
2MASSJ16111330-2019029    & 155  $\pm$  1 &   M3.5  & 0.6 &  17.0    & 0.03 & -1.9 &  0.27  &   9.77$\cdot10^{-10}$  & f  \\
2MASSJ16123916-1859284   & 139  $\pm$  2 &   M1    & 0.6 &   51.0   & 0.22 & -2.3 &  0.50  &   4.75$\cdot10^{-10}$  &  f \\
2MASSJ16133650-2503473    & 145   & M3   & 1.0 &  86.0    & 0.11 & -1.6 &  0.32  &   2.93$\cdot10^{-09}$   &  f  \\
2MASSJ16135434-2320342     & 145   & M4.5 &  0.3 &  52.0    & 0.12 & -2.3 &  0.20  &   1.18$\cdot10^{-09}$   & f   \\
2MASSJ16141107-2305362  & 145   & K4  & 0.3 &  4.0    & 1.05 & -1.4 &  1.25  &   2.09$\cdot10^{-09}$   &  f  \\
2MASSJ16143367-1900133   & 142  $\pm$  2 &   M3   & 1.9 &  69.0   &  0.52 & -2.7 &  0.29  &   5.17$\cdot10^{-10}$  &  f \\
2MASSJ16154416-1921171    & 132  $\pm$  2 &   K7   & 2.8 &   40.0   & 0.30 & -0.3 &  0.81  &   2.44$\cdot10^{-08}$   & f \\
2MASSJ16181904-2028479    & 138  $\pm$  2 &   M5   & 1.6 &   56.0   & 0.05 & -3.4 &  0.16  &   8.05$\cdot10^{-11}$ &  e   \\
\hline 
\end{longtable}
\tablefoot{Disk inclinations from \citet{Barenfeld2017}. Disk types from \citet{barenfeld16,luhman2012,carpenter2006} \\ f: full disk; td: transitional disk; e: evolved disk; b: binary source}

\section{Aligning the NIR and VIS spectra} \label{sect::aligning}
During the data reduction, aligning the wavelength axis of both the NIR and VIS spectra is a crucial step to ensure the correct velocities are measured in the \henir line. We examined the Pa$\delta$ 1004.94~nm and the Li 670.78~nm lines to measure the velocity correction needed for this alignment. First, the radial velocity and heliocentric velocity corrections are applied to the spectra. We then fit the Li line with a 1D Gaussian to measure any shift from its rest wavelength of 670.78~nm. Both the NIR and VIS spectra are shifted so the Li line in the VIS spectra is centred at 0\kms. The majority of our sample (109 sources in total consisting of 73 sources in Lupus, and 36 in Upper Sco) used the velocity shift measured by the Gaussian fits. In nine spectra (all in Lupus), the median Li shift of -0.5\kms is applied to the spectra. In these cases, the Li line is either not detected or the signal-to-noise is too low to properly fit with a Gaussian. 

Next, the Pa$\delta$ 1004.94~nm line, in both the NIR and VIS spectra, is fitted using a 1D Gaussian. The velocity difference between the peak of the Pa$\delta$ lines in the NIR and VIS spectra is used to shift the NIR arm, to be aligned with the VIS arm. For the majority of our spectra, the NIR and VIS arms were already well-aligned, so a total of 73 sources (54 in Lupus and 19 in Upper Sco) are not shifted based on the Pa$\delta$ line.
In 45 sources (28 in Lupus, 17 in Upper Sco), the velocity correction found with the Pa$\delta$ line is applied to align the NIR and VIS spectra. However, in eight (seven in Lupus, one in Upper Sco) of these sources the Pa$\delta$ shift measured with the Gaussian fit was too large ($> \pm$ 20~\kms), and was not appropriate for aligning the spectra. This is due to a low signal-to-noise ratio in the Pa$\delta$ line for these sources. In these cases, the median value of the non-zero Pa$\delta$ shifts was used - this is 9~\kms in Lupus, and -2.9~\kms in Upper Sco.

\begin{figure}[H]
\centering
\includegraphics[width=0.6\textwidth]{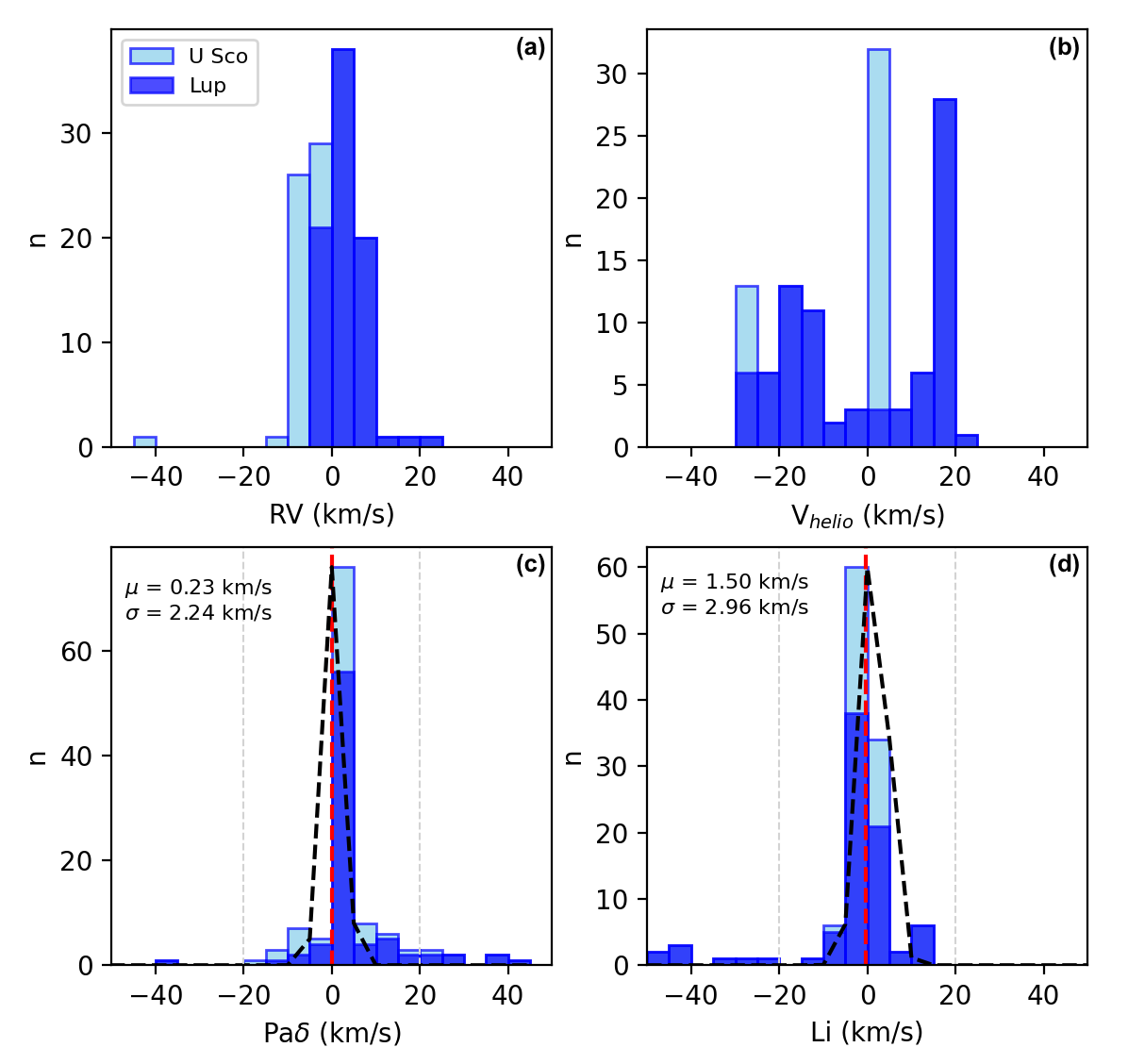}
\caption{Histograms of the velocity corrections applied to the spectra for sources in both star forming regions. Panels (a) and (b) show the radial velocity and the heliocentric velocity corrections. Panels (c) and (d) show the velocity corrections measured from the Pa$\delta$ and Li lines. The red dashed line marks the median value of the Pa$\delta$ and Li shifts, with the corresponding mean and standard deviation listed in each panel.}
\label{fig:vshifts}
\end{figure}

In Figure \ref{fig:vshifts}, we plot histograms of all velocity corrections applied to our spectra in both star forming regions. In each panel, dark blue bars represent sources in Lupus, while light blue shows the Upper Sco sources. Panels (a) and (b) show the radial velocity and the heliocentric velocity corrections. Panels (c) and (d) show the velocity corrections measured from the Pa$\delta$ and Li lines. The red dashed line marks the median value of the Pa$\delta$ and Li shifts, with the corresponding mean and standard deviation listed in each panel. Vertical dashed lines in panels (c) and (d) mark the velocities above which we use the median velocity shifts listed above.
We find the standard deviations for both the Li line and Pa$\delta$ line velocity corrections, $\sigma$, to be less than 3~\kms (2.96~\kms in the Li line velocity shifts and 2.24~\kms for the Pa$\delta$ line). Adding these in quadrature, the combined error due to the velocity corrections is 3.7~\kms and thus do not introduce large uncertainties in the observed maximum velocities in the \henir line (see Section \ref{sect::Res}). Also, since 3$\sigma$ is less than the velocity resolution (of $\approx$ 11~\kms and 16~\kms for the VIS and NIR arms, respectively), these velocity shifts are not likely to affect the number of each profile type observed in the sample.

\section{\henir profiles in the Lupus sample}

\begin{figure}[H]
    \centering
    \includegraphics[width=0.7\textheight]{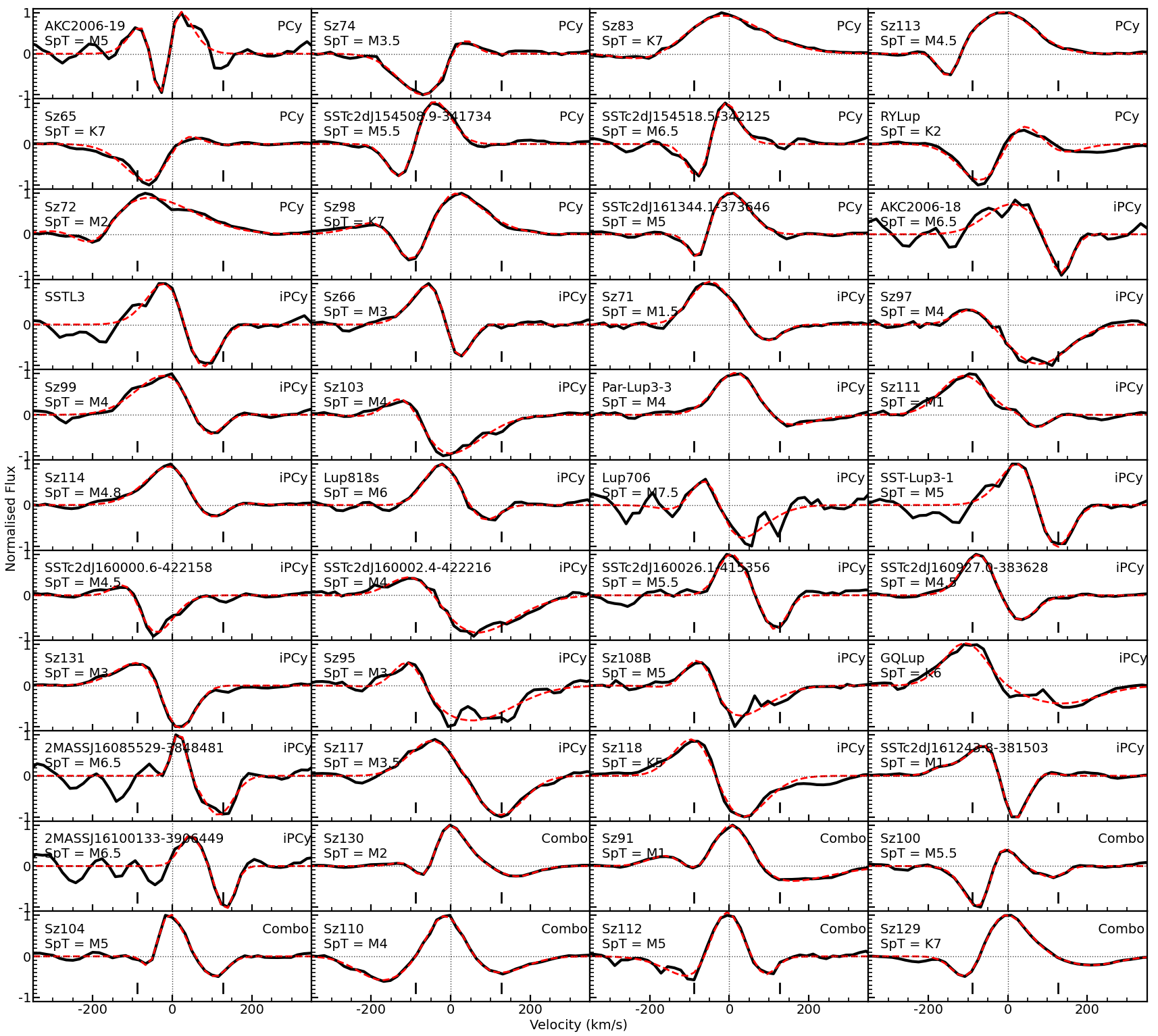}
    \caption{\henir profiles of the Lupus sample, grouped by profile type. The profiles are normalised to the continuum level and rescaled between $\pm$ 1 for plotting. The black vertical dashes (at -88~\kms and +126~\kms) mark photospheric absorption features (i.e Si I 10830.1 \AA and NaI 10837.8 \AA)}
    \label{fig:Lup1}
\end{figure}

\begin{figure}[H]
    \centering
    \includegraphics[width=0.7\textheight]{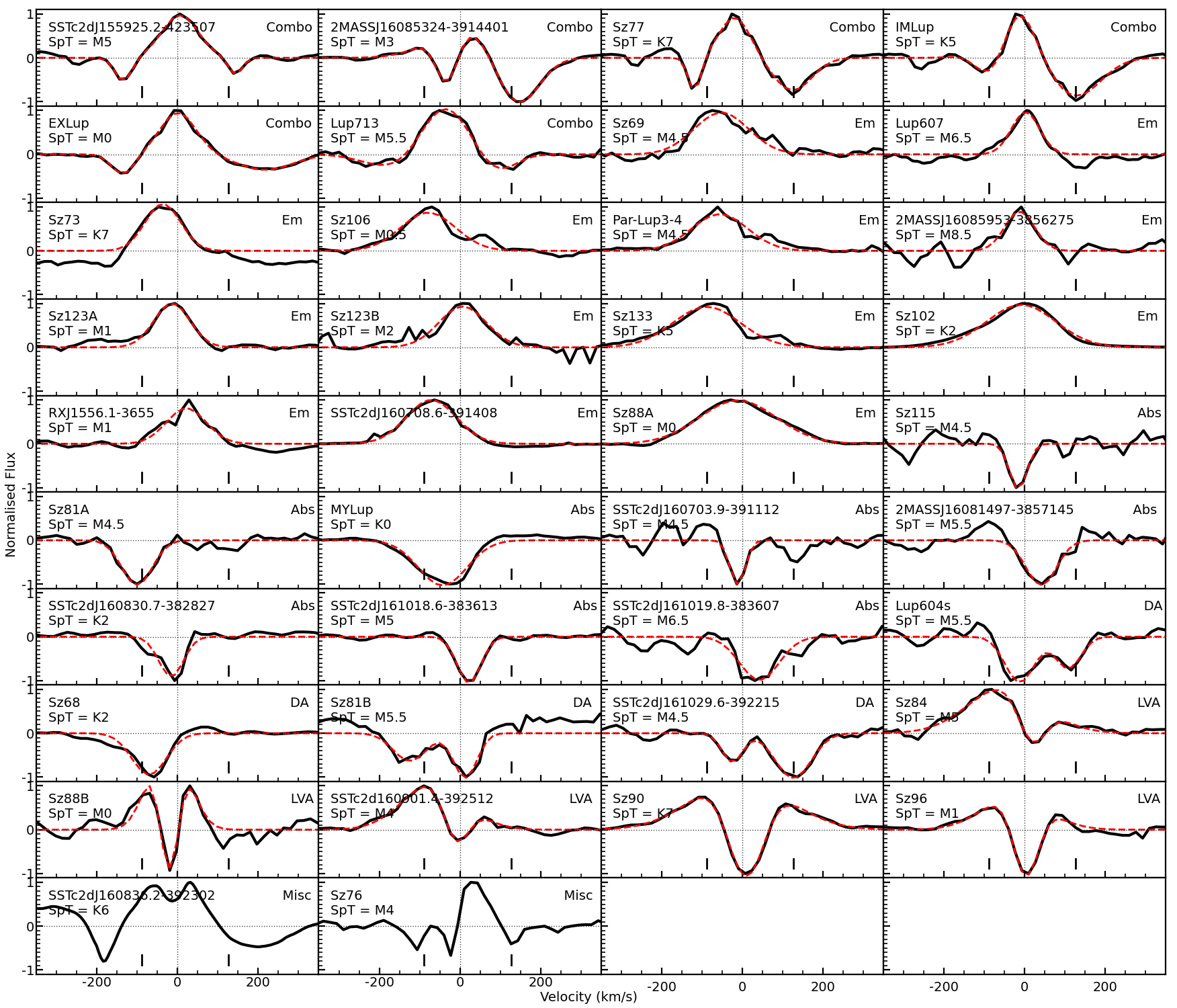}
    \caption{\henir profiles of the Lupus sample, grouped by profile type. The profiles are normalised to the continuum level and rescaled between $\pm$ 1 for plotting. The black vertical dashes (at -88~\kms and +126~\kms) mark photospheric absorption features (i.e Si I 10830.1 \AA and NaI 10837.8 \AA)}
    \label{fig:Lup2}
\end{figure}

We present \henir profiles of the Lupus sample arranged by the fitted profile types in Figures \ref{fig:Lup1} \& \ref{fig:Lup2}. The profiles are normalised to the continuum level and rescaled between $\pm$ 1 for plotting. We mark two photospheric absorption features (i.e Si I 10830.1 \AA and NaI 10837.8 \AA, at -88~\kms and +126~\kms, respectively) on the each plot, as these may contribute to the observed \henir absorption features.

\newpage
\section{\henir profiles in the Upper Scorpius sample}

\begin{figure}[H]
    \centering
    \includegraphics[width=0.7\textheight]{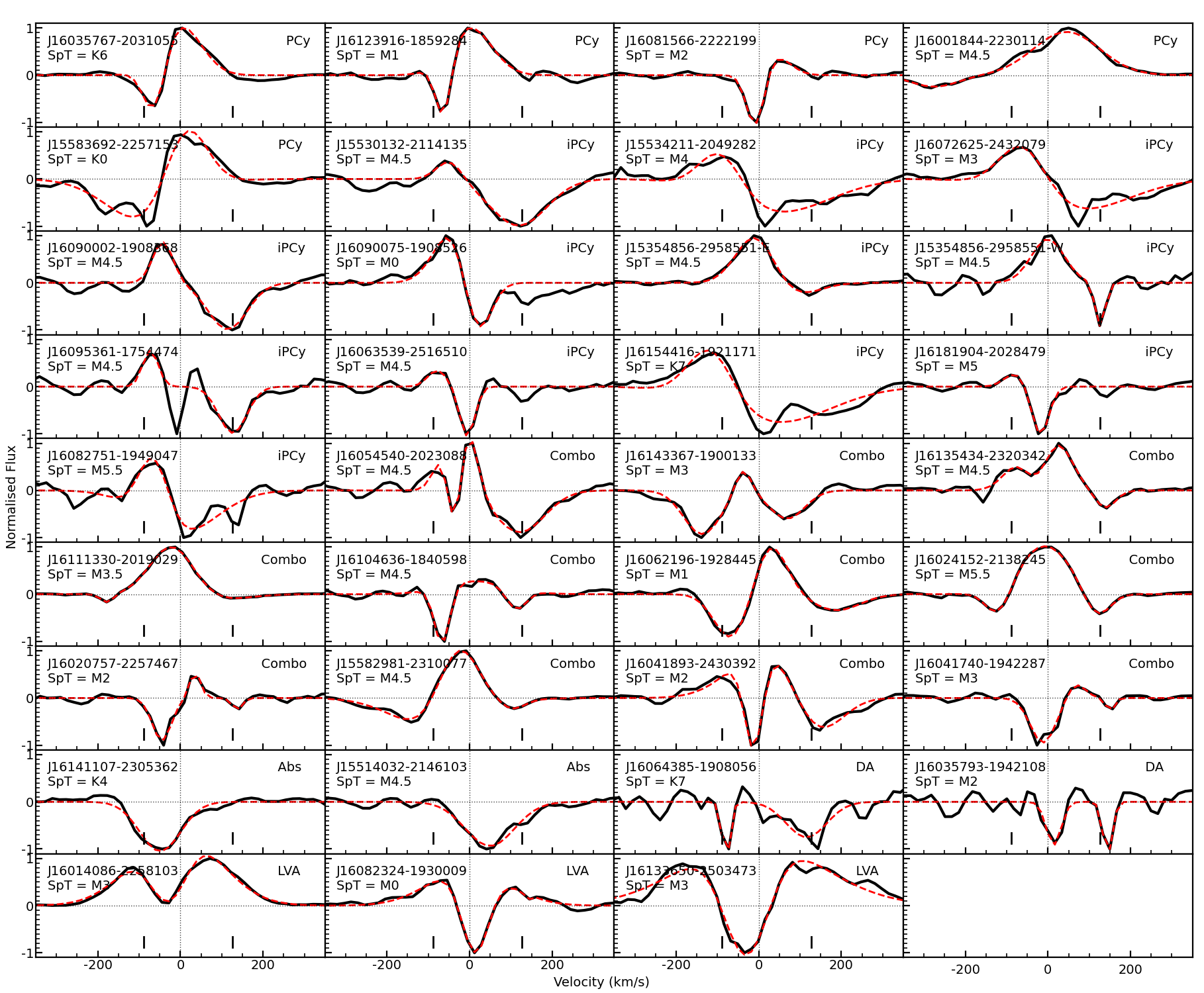}
    \caption{\henir profiles of the Upper Scorpius sample, grouped by profile type. The profiles are normalised to the continuum level and rescaled between $\pm$ 1 for plotting. The black vertical dashes (at -88~\kms and +126~\kms) mark photospheric absorption features (i.e Si I 10830.1 \AA and NaI 10837.8 \AA)}
    \label{fig:USco}
\end{figure}

We present \henir profiles of the Upper Scorpius sample arranged by the fitted profile types in Figure \ref{fig:USco}. The profiles are normalised to the continuum level and rescaled between $\pm$ 1 for plotting. We mark two photospheric absorption features (i.e Si I 10830.1 \AA and NaI 10837.8 \AA, at -88~\kms and +126~\kms, respectively) on the each plot, as these may contribute to the observed \henir absorption features.

\newpage
\section{\henir profiles in Class III sources}
\begin{figure}[h]
    \centering
    \includegraphics{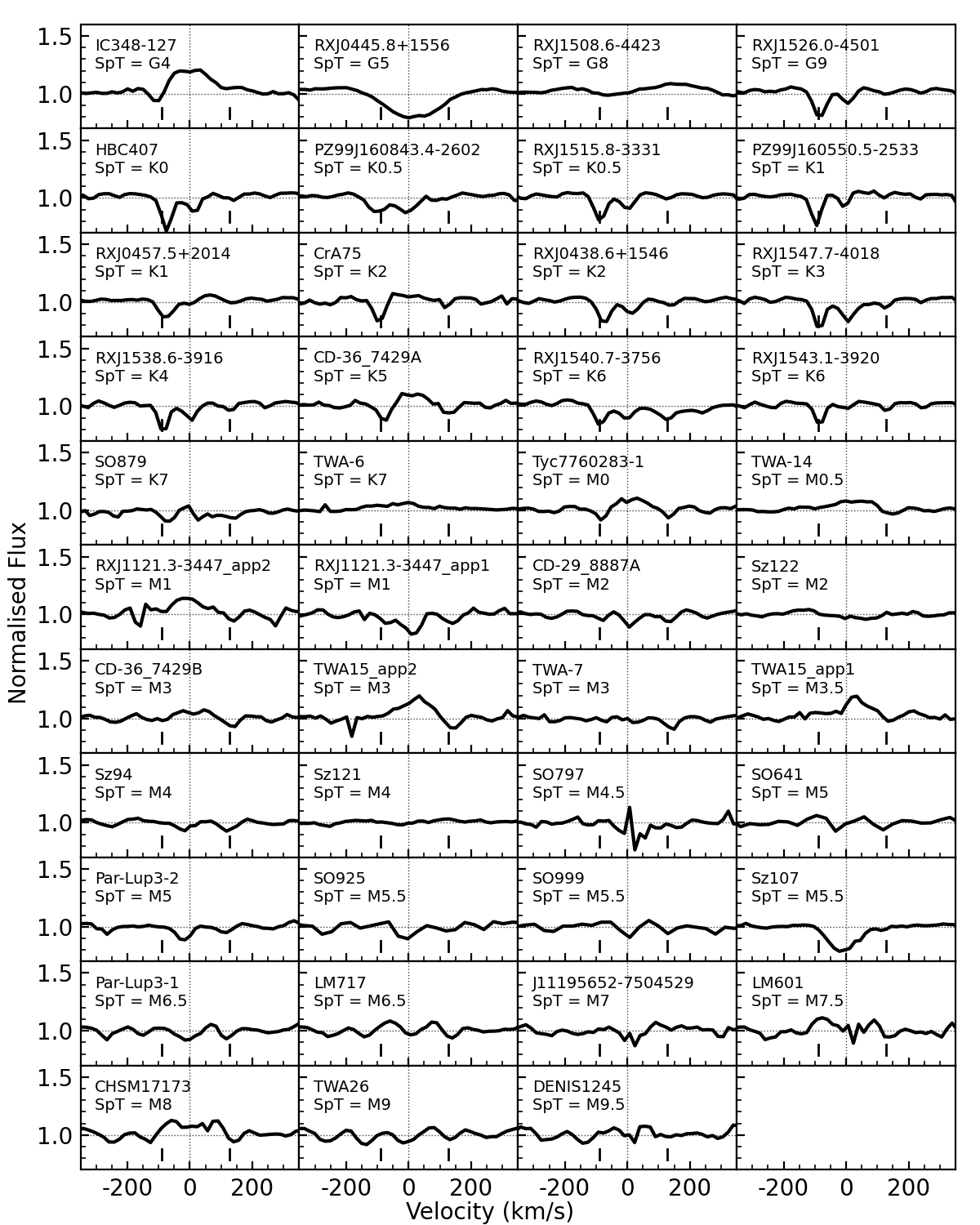}
    \caption{\henir profiles for a sample of 43 Class III sources. We do not observe absorption nor emission features suitable for classifying the profiles as in Section \ref{sect:profile_types}. The black vertical dashes (at -88~\kms and +126~\kms) mark photospheric absorption features (i.e Si I 10830.1 \AA and NaI 10837.8 \AA)}
    \label{fig:classIII}
\end{figure}

43 Class III spectra, from Manara et al. (2013,2017), are plotted centred on the \henir line over a velocity interval of $\pm$ 350~\kms. The subplots are arranged by spectral type (noted on each panel in Figure \ref{fig:classIII}). Each spectrum was normalized to the continuum level by taking the mean continuum level on either side of the He I line. All spectra are corrected for the radial and heliocentric velocities. The grey vertical line marks 0 km/s. The black vertical dashes (at -88~\kms and +126~\kms) mark strong photospheric absorption features (i.e Si I 10830.1 \AA and NaI 10837.8 \AA) which are particularly strong in SpTs G and K. For K-type SpT there also appears to be a persistent absorption feature centred on 0 km/s. Generally, little to no emission is seen for these sources. Strong and broad absorption centred on 0 km/s is seen in only two sources (Sz107 and RXJ0445.8+1556).

\end{appendix}

\end{document}